\documentclass[aps, prb, reprint, showpacs, amsmath, amssymb, dvips]{revtex4-1}

\usepackage{graphicx}

\usepackage{bm}
\newcommand{\drm}{\text{d}}
\newcommand{\im}{\text{i}}
\newcommand{\grad}{\nabla}
\newcommand{\e}{\text{e}}
\newcommand{\del}{\partial}
\newcommand{\crm}{\text{c}}

\newcommand{\srm}{\text{s}}
\newcommand{\rbm}{\bm{r}}

\newcommand{\Abm}{\bm{A}}
\newcommand{\Bbm}{\bm{B}}
\newcommand{\Hbm}{\bm{H}}

\newcommand{\Fcal}{\mathcal{F}}
\newcommand{\TcSB}{T_\text{c}^\text{(s)}}
\newcommand{\TcTB}{T_\text{c}}
\newcommand{\Tc}{T_{\text{c} 0}}
\newcommand{\TczSB}{T_\text{c0}^\text{(s)}}
\newcommand{\Tcbar}{\overline{T}_{\text{c} 0}}
\newcommand{\bbar}{\overline{b}}
\newcommand{\Nbar}{\overline{N}}
\newcommand{\qbar}{\overline{q}}
\newcommand{\kbar}{\overline{k}}
\newcommand{\mubar}{\overline{\mu}}
\newcommand{\epsbar}{\overline{\epsilon}}
\newcommand{\Ebar}{\overline{E}}
\newcommand{\ktil}{\widetilde{k}}
\newcommand{\gamtil}{\widetilde{\gamma}}
\newcommand{\etatil}{\widetilde{\eta}}
\newcommand{\no}{\nonumber}

\begin{document}
\title{Fluctuation Diamagnetism in Two-Band Superconductors}
\author{Kyosuke Adachi}
\author{Ryusuke Ikeda}
\affiliation{Department of Physics, Kyoto University, Kyoto 606-8502, Japan}
\date{\today}
\begin{abstract}
Anomalously large fluctuation diamagnetism around the superconducting critical temperature has been recently observed on iron selenide (FeSe) [S. Kasahara \textit{et al}., unpublished]. 
This indicates that superconducting fluctuations (SCFs) play a more significant role in FeSe, which supposedly has \textit{two}-\textit{band} structure, than in the familiar \textit{single}-\textit{band} superconductors. 
Motivated by the data in FeSe, SCF-induced diamagnetism is examined in a two-band system, on the basis of a phenomenological approach with a Ginzburg-Landau functional. 
The obtained results indicate that the SCF-induced diamagnetism may be more enhanced than that in a single-band system due to the existence of two distinct fluctuation modes. 
Such enhancement of diamagnetism unique to a two-band system seems consistent with the large diamagnetism observed on FeSe, though still far from a quantitative agreement. 
\end{abstract}
\maketitle

%%%%%%%%%%%%%%%%%

\section{Introduction}

Near the superconducting critical temperature $T_\crm$, the thermal fluctuation can elicit electron pairs over a finite range of space and time even in the normal phase above $T_\crm$. 
Such transient pairs, referred to as superconducting fluctuation (SCF), can induce precursory phenomena of superconductivity; 
For example, it suppresses resistivity and enhances diamagnetism. 
SCF-induced phenomena have been investigated and observed for nearly half a century.\cite{Skocpol, Tinkham, Larkin} The corresponding fluctuation effect in Fermi-superfluids has also been seen in unitary Fermi gas systems.\cite{Ku} 
Importance of thermal fluctuation is usually signaled by the width of the so-called critical region, which is larger in a superconductor with a higher $T_\crm$ and/or a shorter coherence length.\cite{Larkin} 
In this sense, high-temperature cuprate superconductors\cite{cuprate} have been preferable for studying SCF. 
They have accelerated understanding of SCFs especially in a high magnetic field, where the Gaussian approximation neglecting the mode coupling (mutual interaction between the fluctuations) 
becomes useful only at much higher temperatures than in zero-field case due to the effective reduction of the dimensionality of the fluctuation.\cite{Ikeda}
In high magnetic fields, the resulting features generated by SCF emerge, such as the crossing of magnetization curves\cite{Rosenstein, Tesanovic, Jiang, PJLin} and the scaling law of both thermodynamic quantities and transport coefficients.\cite{Welp, Ikeda, Tesanovic, Ullah} 

To describe the SCF theoretically, one often needs to construct the Ginzburg-Landau (GL) free-energy functional based on a microscopic (electronic) Hamiltonian. As far as effects of SCF on electronic quantities are not considered, however, the SCF effects can be examined starting by introducing a phenomenological GL functional.\cite{Larkin} 
On the other hand, to describe SCF effects in superconductors with moderately strong SCF under a nonzero magnetic field, the mode coupling, i.e., the interaction between SCFs, plays essential roles.\cite{Ikeda} This mode coupling affects the SCF in a couple of different manners. First, the SCF with higher energy pushes the \textit{bare} transition point to a lower temperature through the mode coupling. Further, the SCF with lower energy interacting through the mode coupling determines the crossover behavior in the so-called vortex-liquid regime and even the genuine transition, i.e., the vortex-lattice-melting transition. If one focuses on the vortex-liquid regime and the region at higher temperatures, a convenient treatment for describing the SCF interaction is the so-called Hartree\cite{Ikeda, Tesanovic, Ullah, Welp, Rosenstein, PJLin} or Hartree-Fock (HF) approximations.\cite{Jiang} 

Among the superconductors which have been examined recently, iron selenide (FeSe), which is one of typical iron-based superconductors has been found to show unusually remarkable fluctuation effects. Although SCF-induced phenomena have also been discussed previously in other iron-based superconductors,\cite{Ramos, Bossoni, Prando} 
%The recently discovered iron-based superconductors have been examined mainly co%ncerning their electronic structure, magnetic order, and pairing symmetry.\cite%{Paglione} 
very recent magnetic-torque measurement on iron selenide (FeSe) indicates the presence of anomalously large SCF-induced diamagnetism even far above the transition temperature.\cite{Kasahara} 
In addition, a crossing behavior of magnetization curves at different magnetic fields, which is a typical SCF phenomenon seen repeatedly in 2D-like high $T_\crm$ cuprates, has been observed in this 3D-like superconductor in high magnetic fields.\cite{Kasahara} Since FeSe (as well as similar compounds\cite{Lubashevsky, Okazaki}) is supposed to have very small Fermi surfaces,\cite{Kasahara2, Terashima} the zero temperature coherence length is estimated to be comparable with the averaged inter-electron distance. 
It thus seems plausible that the thermal fluctuation is important in FeSe. 
However, the observed diamagnetism is much larger than that expected from a conventional \textit{single}-\textit{band} theory. On the other hand, 
it should be noted that several experiments have suggested \textit{two}-\textit{band} structure of FeSe.\cite{Terashima, Kasahara2} 
Therefore, it is natural to theoretically examine whether or not a two-band character can explain the anomalously large SCF-induced diamagnetism observed in FeSe. 

Motivated by the experiment on FeSe mentioned above, we examine the SCF-induced diamagnetism in a two-band system {\it both} in low and high magnetic fields. 
Starting with a GL functional, the non-Gaussian SCF including the mode-coupling effects will be examined within the HF approximation. 
As a result, we find that the SCF-induced diamagnetism is more enhanced in a two-band system than in a single-band system. 
This enhancement is caused by a character of a two-band system, as discussed in the following sections. 

This paper is organized as follows. 
In Sec. \ref{sec:MM}, our model and calculation method are explained. 
In Sec. \ref{sec:results}, we explain our calculation results and consider a two-band character, and then, a comparison between our results and the experimental data on FeSe\cite{Kasahara} will be done. 
In Sec. \ref{sec:summary}, this study is summarized with conclusions. 

%%%%%%%%%%%%%%%%%

\section{Model and Method}
\label{sec:MM}

In this section, we give a detailed account of the way of investigating the SCF-induced diamagnetism. Following explanation of our model (Sec. \ref{sec:model}), we first discuss the case of a single-band system (Sec. \ref{sec:methodSB}) and then apply our method to a two-band system (Sec. \ref{sec:methodTB}). 
Note that some quantities below are represented explicitly in Appendices \ref{app:exp} and \ref{app:expTB}. 

\subsection{Model}
\label{sec:model}

To examine a character of diamagnetism in a two-band system with volume $V = L_x L_y L_z$ under an external magnetic field $H$, we choose a symmetric two-band GL functional as a starting point: 
\begin{eqnarray}
\Fcal &=& \int_V \drm^3 \rbm \, \alpha \bigg[ \epsilon^{(\srm)} \left( |\psi_1|^2 + |\psi_2|^2 \right) 
+ \eta \left( \psi_1^* \psi_2 + \psi_2^* \psi_1 \right) 
\nonumber \\
&& \hspace{45pt} 
+ ( \xi \bm{\Pi} \psi_1 )^\dag ( \xi \bm{\Pi} \psi_1 )
\nonumber \\
&& \hspace{45pt} 
+ ( \xi \bm{\Pi} \psi_2 )^\dag ( \xi \bm{\Pi} \psi_2 ) 
\nonumber \\
&& \hspace{45pt} 
+ \frac{\beta}{2 \alpha} \left( |\psi_1|^4 + |\psi_2|^4 \right) \bigg]. 
\label{eq:Fcal}
\end{eqnarray}
Here $\psi_n (\rbm)$ ($n = 1, 2$) is a fluctuating order-parameter field with a band index $n$, $\eta$ is an interband coupling, which commonly appears in any two-band system, and $\epsilon^{(\srm)} = \ln (T / T^{(\srm)}_{\crm 0})$ is a reduced temperature with the bare, i.e., unrenormalized, critical temperature in $\eta = 0$ limit $T^{(\srm)}_{\crm 0}$. 
Here, the upper index $(\srm)$, implying $\eta=0$, in this mass term is introduced to avoid any confusion with the corresponding two mass terms in the two-band case to be discussed later. 
Further, $\bm{\Pi} = - \im \grad + (2\pi/\phi_0)\Abm$ is the gauge-invariant momentum operator with the flux quantum $\phi_0$ and a vector potential $\Abm (\rbm)$, and $\xi \equiv \xi_{ij}$ is a matrix of the GL coherence length when $\eta = 0$; $\xi_{ij} = \xi_{ab} \delta_{ij} (1 - \delta_{iz}) + \xi_{c} \delta_{iz} \delta_{jz}$. 
Regarding the coherence lengths, we have not attached the index $(\srm)$ to them because it is found that, in contrast to the mass term, a nonvanishing $\eta$ does not lead to any change of the coherence lengths in the symmetric two-band case on which we focus in the present work (see below). 
Note that we only have to set $\psi_2 \equiv 0$ if we consider a single-band case. 

In this model, the system is assumed to be a strongly type II superconductor (the GL parameter $\kappa \gg 1$). 
Thus the contribution of supercurrent to the forth-order term\cite{Ruggeri} is negligible, and the induction is nearly equal to the external field ($\Bbm \simeq \mu_0 \Hbm$) as long as the effect of diamagnetism is small. 
This approximation is appropriate to describing the iron-based superconductors. 
In particular, as far as the fluctuation effect is concerned, taking the type-II limit is safely valid. 

For simplicity, we consider the case where the external field is parallel to the $c$ axis ($\Hbm \parallel c$) by using the Landau gauge $\Abm(\rbm) = (0,Bx,0)$. When $\Hbm \parallel ab$, we only have to replace $\xi_{ab}$ and $\xi_c$ with $\sqrt{\xi_{ab} \xi_c}$ and $\xi_{ab}$, respectively in the final expression in $\Hbm \parallel c$ case.\cite{Hdirection} 

Our symmetric model might be too simple to capture the detailed feature of the two-band structure of FeSe. \cite{Terashima,Kasahara2} Nevertheless, we believe that main results on the fluctuation effects of a general asymmetric two-band model will be captured more clearly by studying the symmetric model with a smaller number of phenomenological parameters. 

\subsection{Single-Band Case}
\label{sec:methodSB}

For completeness, we begin with the case of a single-band system. 
A single-band limit ($\psi_2 \equiv 0$) of $\Fcal$ [Eq. (\ref{eq:Fcal})] leads to a GL functional describing this system: 
\begin{eqnarray}
\Fcal^{(\srm)} &=& \int_V \drm^3 \rbm \, \alpha \bigg[ \epsilon^{(\srm)} |\psi|^2
+ ( \xi \bm{\Pi} \psi )^\dag ( \xi \bm{\Pi} \psi ) \nonumber \\
&& \hspace{45pt} + \frac{\beta}{2 \alpha} |\psi|^4 \bigg]. 
\end{eqnarray}
Note that the lower index of the order parameter $\psi_1$ is omitted here. 
To evaluate the diamagnetism, we approximately calculate the fluctuation-induced free-energy density, and then differentiate it with respect to the external field. 
Thus we explain how to estimate the free-energy density as a function of temperature and magnetic field. 
We remark that our treatment of the fluctuation renormalization is, roughly speaking, based on the HF approximation,\cite{Jiang} combined with the fluctuation-energy-cutoff condition.\cite{Carballeira} 

First, we expand the order-parameter field via the Landau levels (LLs). 
Each eigenfunction of LLs is specified by the LL index $N$ and wavevector $(q, k)$. 
The order parameter is then expressed as 
\begin{equation}
\psi (\rbm)
= \sum_{N, q, k} \varphi_{N q} (x) \frac{\e^{\im q y}}{\sqrt{L_y}} \frac{\e^{\im k z}}{\sqrt{L_z}} \, a_{N q k}. 
\label{eq:OPexpSB}
\end{equation}
Here, $\varphi_{N q} (x)$ is the normalized eigenfunction in the $N$th LL with a label $q$, which measures the LL degeneracy, and $a_{N q k}$ is a dimensionless fluctuating amplitude for each mode. 
Next, we diagonalize the second-order term of $\Fcal^{(\srm)}$ in $a_{Nqk}$ as
\begin{eqnarray}
\Fcal^{(\srm)}_\text{2nd}
&\equiv& \int_V \drm^3 \rbm \, \alpha \left[ \epsilon^{(\srm)} |\psi|^2
+ ( \xi \bm{\Pi} \psi )^\dag ( \xi \bm{\Pi} \psi ) \right] \nonumber \\
&=& \sum_{N, q, k} \alpha \left[ \epsilon^{(\srm)} + 2 h \left( N + \frac{1}{2} \right) + {\xi_c}^2 k^2 \right] |a_{N q k}|^2, 
\hspace{10pt}
\label{eq:F2ndSB}
\end{eqnarray}
where $h = 2 \pi {\xi_{ab}}^2 \mu_{0} H / \phi_0$ is a dimensionless magnetic field. 
In the single-band case discussed here, we also define a dimensionless temperature as $t = T / \Tc^{(\srm)}$. 

First, the Gaussian approximation will be discussed concentrating on temperatures far above the zero field critical region. 
In this case, we can neglect the mode-coupling effects due to the forth-order term of $\Fcal^{(\srm)}$ (see below). 
As the temperature increases away from the critical region, the contribution of short-wavelength fluctuations, i.e., fluctuations with a wavelength as well as or shorter than the GL coherence length, becomes nonnegligible. 
However, a GL functional, such as our model $\Fcal$ or $\Fcal^{(\srm)}$, usually overestimates them.\cite{Skocpol} 
In evaluating not only the free energy but the magnetization, we need to introduce a certain cutoff that effectively restricts the short-wavelength fluctuations.\cite{com1}

Among the cutoff schemes used previously, introducing an energy cutoff is known to lead to a consistent description of the behavior of short-wavelength fluctuations far beyond the critical region in low fields.\cite{Carballeira} 
Following Ref. \onlinecite{Carballeira}, we introduce the upper limit of the LL index $N$ and that of the wavenumber $k$: 
\begin{equation}
N_\text{cut}  = \frac{c - \epsilon^{(\srm)}}{2h} - 1, \ 
k_\text{cut} = \frac{\sqrt{c-\epsilon^{(\srm)}}}{\xi_c}, 
\label{eq:cutoffSB}
\end{equation}
where $c = \mathcal{O}(1)$ is a dimensionless cutoff parameter. 
In the present work, these cutoff conditions will be used to discuss the phenomena in high fields where $h \gtrsim 0.1$. 

We proceed to the effects of the mode coupling due to the last term of $\Fcal^{(\srm)}$. 
In zero field, the mode coupling affects the SCF behaviors in two ways. First, the mode coupling pushes the bare transition temperature $\Tc^{(\srm)}$ down to a lower temperature $T_\crm^{(\srm)}(0)$, at which the long-ranged phase coherence sets in. Second, it induces a critical region around the genuine transition temperature $T_\crm^{(\srm)}(0)$ in which a universal critical behavior is seen. In a nonzero magnetic field, high energy SCF shifts the bare depairing field $\phi_0 |\epsilon^{(\srm)}|/(2 \pi {\xi_{ab}}^2)$ to a renormalized depairing field $H_\text{c2}^{(\srm)}(T)$ lying at a lower field,\cite{RI95} and no genuine transition occurs at this $H_\text{c2}^{(\srm)} (T)$. Consistently, the onset of the phase coherence is pushed down to zero temperature,\cite{Moore,RI92} and the genuine superconducting transition becomes the first-order vortex-lattice-melting transition occurring below $H_\text{c2}^{(\srm)}$. The non-Gaussian-SCF-dominated region between $H_\text{c2}^{(\srm)}$ and the melting-transition line is often called as the vortex-liquid regime, which, broadly speaking, corresponds to the critical region in zero field. Hereafter, the mode-coupling effects will be examined by focusing on the diamagnetism below and above $H_\text{c2}^{(\srm)}$. In the following, we refer to $H_\text{c2}^{(\srm)} (T)$ as $T_\text{c}^{(\srm)} (H)$ when it is expressed as a temperature dependent on $H$. 

To investigate the diamagnetic SCF behaviors, we treat the mode-coupling effects by using a variational method equivalent to the HF approximation. 
Compared with the familiar Gaussian approximation,\cite{Skocpol} this method enables us to consistently treat a wider temperature and field range in the vortex-liquid region at least at a qualitative level. 
This variational method will be sketched in the following. Details of this method are explained in Appendix. 
Using a variational parameter $\mu$, we divide the GL functional into two parts: $\Fcal^{(\srm)} = \Fcal_0^{(\srm)} + \Fcal_1^{(\srm)}$, where
\begin{equation}
\Fcal^{(\srm)}_0 = \sum_{N,q,k} \alpha \left[ \mu + 2h \left( N + \frac{1}{2} \right) + {\xi_c}^2 k^2 \right] |a_{Nqk}|^2. 
\end{equation}
Here $\mu$ can be interpreted as a renormalized one of $\epsilon^{(\srm)}$. 
Then we obtain a trial free-energy density $f^{(\srm)}_\text{tri} (t, h; \mu)$ that is always larger than the exact free-energy density $f^{(\srm)}_\text{ex} (t, h)$ as follows: 
\begin{eqnarray}
f^{(\srm)}_\text{ex} 
&=& -(T / V) \ln \text{Tr}_{a} \, \e^{-\Fcal^{(\srm)} / T} \nonumber \\
&=& -(T / V) \ln \text{Tr}_{a} \, \e^{-\Fcal^{(\srm)}_0 / T}
 -(T / V) \ln \langle \e^{-\Fcal^{(\srm)}_1 / T} \rangle_0 \nonumber \\ &\leq& f^{(\srm)}_0 + \frac{\langle \Fcal^{(\srm)}_1 \rangle_0}{V} \equiv f^{(\srm)}_\text{tri}, 
\end{eqnarray}
where $\langle \cdots \rangle_0 = \text{Tr}_a \, ( \cdots ) \, \e^{-\Fcal^{(\srm)}_0 / T} / \, \text{Tr}_a \, \e^{-\Fcal^{(\srm)}_0 / T}$ is the canonical average with respect to $\Fcal^{(\srm)}_0$, and $\text{Tr}_a ( \cdots )$ corresponds to integrating all degrees of freedom of the order-parameter fields with the above cutoff conditions [Eq. (\ref{eq:cutoffSB})]: 
\begin{eqnarray}
&& \text{Tr}_a ( \cdots ) \nonumber \\
&& = \int_{- \infty}^{\infty} \prod_{N = 0}^{N_\text{cut}} 
\prod_{k = -k_\text{cut}}^{k_\text{cut}} 
\prod_{q} 
\drm \text{Re} \, a_{N q k} \, \drm \text{Im} \, a_{N q k} \, ( \cdots ). \nonumber 
\end{eqnarray}
Here the number of possible $q$ is $\mu_0 H L_x L_y / \phi_0$, which is equal to the degree of degeneracy of each LL. 
Note also that $f^{(\srm)}_0 = -(T / V) \ln \text{Tr}_{a} \, \e^{-\Fcal^{(\srm)}_0 / T}$ is the Gaussian contribution of the renormalized fluctuation to the whole free-energy density, and that $\langle \Fcal^{(\srm)}_1 \rangle_0 / V$ is the residual contribution. 
The former [$f_0^{(\srm)} (t, h; \mu)$] can be reduced to the following form as described in Appendix \ref{app:exp}: 
\begin{equation}
f^{(\srm)}_0
= \frac{T_{\crm 0}^{(\srm)}}{2 \pi^2 {\xi_{ab}}^2 \xi_c} \, t h \,
I_1 (t, h; \epsilon^{(\srm)}; \mu), 
\label{eq:f0SB}
\end{equation}
where $I_1 (t, h; \epsilon^{(\srm)}; \mu)$ is given in Eq. (\ref{eq:I1SB}) in the Appendix \ref{app:exp}. 
The latter ($\langle \Fcal_1^{(\srm)} \rangle_0 / V$), on the other hand, can be written with $f_0^{(\srm)}$ as shown in Appendix \ref{app:fres}. 
As a consequence, we obtain 
\begin{equation}
f^{(\srm)}_\text{tri}
= f^{(\srm)}_0
-(\mu - \epsilon^{(\srm)}) \frac{\del f^{(\srm)}_0}{\del \mu}
+ \frac{\beta}{\alpha^2} \left( \frac{\del f^{(\srm)}_0}{\del \mu} \right)^2. 
\label{eq:ftriSB}
\end{equation}
We get the optimal free-energy density $f^{(\srm)}_\text{opt} (t, h)$ by minimizing $f^{(\srm)}_\text{tri} (t, h; \mu)$ with respect to $\mu$. 
The optimizing equation $\del f^{(\srm)}_\text{tri} / \del \mu = 0$ leads to the equation for $\mu$: 
\begin{equation}
\mu = \epsilon^{(\srm)} + \frac{2 \beta}{\alpha^2} \frac{\del f^{(\srm)}_0}{\del \mu}
\label{eq:OPeqSB}
\end{equation}
We refer to $\mu$ that satisfies this equation as $\mu (t, h)$ hereafter. 

The magnetization is calculated from $M^{(\srm)}_\text{dia} = - \del f^{(\srm)}_\text{opt} / \del (\mu_0 H) = -(2 \pi {\xi_{ab}}^2 / \phi_0) \del f^{(\srm)}_\text{opt} / \del h$. 
This formula is simplified as follows: 
\begin{eqnarray}
M^{(\srm)}_\text{dia} (t, h) 
&=& - \frac{2 \pi {\xi_{ab}}^2}{\phi_0} 
\frac{\del f^{(\srm)}_\text{tri} (t, h; \mu)}{\del h} \bigg|_{\mu = \mu (t, h)} \nonumber \\
&=& - \frac{2 \pi {\xi_{ab}}^2}{\phi_0} 
\frac{\del f^{(\srm)}_0 (t, h; \mu)}{\del h} \bigg|_{\mu = \mu (t, h)}. 
\label{eq:MSB}
\end{eqnarray}
In the first equality, we use the fact that $f^{(\srm)}_\text{opt} (t, h) = f^{(\srm)}_\text{tri} [t, h; \mu (t, h)]$ is minimized with respect to $\mu$. 
In obtaining the second line, Eqs. (\ref{eq:ftriSB}) and (\ref{eq:OPeqSB}) are used. 
By combining Eqs. (\ref{eq:OPeqSB}) and (\ref{eq:MSB}) with the explicit form of $f^{(\srm)}_0$, we obtain the magnetization $M^{(\srm)}_\text{dia}$ as a function of dimensionless temperature $t$ and magnetic field $h$. 
The detailed expression of $M_\text{dia}^{(\srm)} (t, h)$ is summarized in Appendix \ref{app:exp}. 

Since we have introduced the cutoff parameter $c$ as indicated in Eq. (\ref{eq:cutoffSB}), both the resultant optimizing equation [Eq. (\ref{eq:OPeqSB}) or (\ref{eq:OPeqSBexp})] and magnetization formula [Eq. (\ref{eq:MSB}) or (\ref{eq:MdiaSBexp})] explicitly depend on $c$. 
Here, we point out that the present results are not in conflict with both of previous works neglecting and including the mode-coupling effect. 
First, we note that we can reproduce the Gaussian magnetization formula\cite{Carballeira} equivalent to that provided by Prange in Ref. \onlinecite{Prange} if we simply neglect the mode-coupling effect and push the cutoff $c$ to infinity (see Appendix \ref{app:cindepSB}). 
Second, in the case where the mode-coupling effect is important, the reader may wonder if the influence of $c$ can be totally absorbed into the renormalized zero-field critical temperature $T_\text{c}^\text{(s)} (0)$ as in the standard approach of the renormalization group.\cite{Amit} 
Indeed, if we replace the mass parameter ($\epsilon^\text{(s)} = \ln (T / \Tc^\text{(s)})$ in this study) with $T / \Tc^\text{(s)} - 1$ following Ref. \onlinecite{Jiang}, we can formally take $c \rightarrow \infty$ limit and obtain the cutoff-independent optimizing equation and magnetization formula with not $\Tc^\text{(s)}$ but $\TcSB (0)$ as the characteristic temperature scale (see Appendix \ref{app:cindepSB}). 
In our formulation, however, a finite cutoff is introduced [$c = \mathcal{O} (1)$] so that we can consistently describe the short-wavelength fluctuations\cite{Carballeira} as mentioned above. 

Before ending this section, we stress that our formalism can treat a broad temperature and field range \textit{in the same framework}. 
First, the cutoff condition [Eq. (\ref{eq:cutoffSB})] enables us to consistently evaluate short-wavelength fluctuations and to reproduce magnetization curves which are also reliable in low fields ($h \ll 1$) and outside the zero field critical region.\cite{Carballeira} 
Second, near and below $H_\text{c2}^{(\srm)}$ in relatively high fields ($h \gtrsim 0.1$), the fluctuation renormalization [Eq. (\ref{eq:OPeqSB})] results in a characteristic behavior of magnetization (see Sec. \ref{sec:results}) commonly seen in the high $T_\crm$ cuprates with strong 
fluctuation.\cite{Li, Welp, Sugui} Previously, 
explanation of these properties has been theoretically tried frequently based on a GL model taking account only of the lowest-LL ($N = 0$) modes \cite{Tesanovic, Rosenstein, PJLin} in contrast to the present model also taking account of the contribution from high LLs. As is shown in Appendix D, however, it is not easy, as far as the magnetization is concerned, to justify the approach focusing only on the lowest-LL fluctuation modes. \cite{Oregon} 
Lastly, we briefly comment on a formalism developed in Ref. \onlinecite{Jiang}, where the authors also take account of high LLs and use a similar method to the variational method used here. 
They consider a quasi-two-dimensional system, which is different from an anisotropic three-dimensional system of our interest. 
In addition, though they also introduce a cutoff of the LL index, their method seems different from ours [Eq. (\ref{eq:cutoffSB})]. The latter consistently describes the short-wavelength fluctuations even in low enough fields.\cite{Carballeira} 
In summary, the present variational method seems qualitatively appropriate both in low and high fields as well as both close to and above the renormalized depairing field $H_\text{c2}^{(\srm)}$. 

\subsection{Two-Band Case}
\label{sec:methodTB}

In this section, we extend our formalism illustrated in Sec. \ref{sec:methodSB} to a two-band system described by the original GL functional [Eq. (\ref{eq:Fcal})]. 
The basic idea is the same as that of the single-band case; 
Our treatment of the fluctuation renormalization is based on the HF approximation, joined together with the fluctuation-energy-cutoff condition. 

As a preliminary, we mention the effects of the interband coupling (represented as $\eta$ in our model) on the fluctuation properties. 
The system on which we focus here is symmetric with respect to the exchange between the two bands. 
The bare critical temperature $\Tc^{(\srm)}$ is thus common to both bands as long as the two bands are independent, i.e., $\eta = 0$. 
If $\eta \neq 0$, however, this degeneracy is resolved by the mixing of fluctuations in each band and consequently, a finite interband coupling creates two distinct modes: 
One mode with lower excitation energy has a higher bare critical temperature $\Tc$, and the other higher energy mode has a lower bare critical temperature $\Tcbar$. We refer to the former as the low-energy mode (LEM), and the latter as the high-energy mode (HEM). 
In the following, we consider the general case where $\eta \neq 0$. 
Just like in the single-band case, we refer to the genuine transition temperature in zero field as $T_\crm (0)$ and to the temperature corresponding to $H_\text{c2} (T)$ as $T_\crm (H)$.

To specifically see how the LEM and HEM appear, we diagonalize the second-order term of $\Fcal$. 
First, we expand the order-parameter fields via the LLs as in the single-band case: 
\begin{equation}
\psi_n (\rbm) = \sum_{N,q,k} \varphi_{Nq} (x) \frac{\e^{\im q y}}{\sqrt{L_y}} \frac{\e^{\im k z}}{\sqrt{L_z}} 
a_{nNqk}, 
\label{eq:OPexpTB}
\end{equation}
where $\varphi_{Nq}(x)$ is the normalized eigenfunction in the $N$th LL with a label $q$ as in the single-band case. 
This expansion leads to
\begin{eqnarray}
&& \Fcal_\text{2nd} \nonumber \\
&& \equiv \int_V \drm^3 \rbm \, \alpha \big[ \epsilon^{(\srm)} \left( |\psi_1|^2 + |\psi_2|^2 \right) 
+ \eta \left( \psi_1^* \psi_2 + \psi_2^* \psi_1 \right) 
\nonumber \\
&& \hspace{55pt} 
+ ( \xi \bm{\Pi} \psi_1 )^\dag ( \xi \bm{\Pi} \psi_1 )
\nonumber \\
&& \hspace{55pt} 
+ ( \xi \bm{\Pi} \psi_2 )^\dag ( \xi \bm{\Pi} \psi_2 ) \big]
\nonumber \\
&& = \sum_{N,q,k} \alpha 
\bigg\{
\left[ \epsilon^{(\srm)} + 2 h \left( N + \frac{1}{2} \right) + {\xi_{c}}^2 k^2 \right] \nonumber \\
&& \hspace{85pt} \times \left( |a_{1Nqk}|^2 + |a_{2Nqk}|^2 \right) \nonumber \\
&& \hspace{50pt} + \eta \left( a_{1Nqk}^* a_{2Nqk} + a_{2Nqk}^* a_{1Nqk} \right)
\bigg\}.
\label{eq:F2nd}
\end{eqnarray}
Here, we introduce a dimensionless magnetic field as $h = 2 \pi {\xi_{ab}}^2 \mu_0 H/\phi_0$. 
Note that $a_{nNqk}$ represents a dimensionless fluctuating amplitude of the $n$th band. Then, we now only have to diagonalize Eq. (\ref{eq:F2nd}) in terms of a $2 \times 2$ matrix $M_{Nk}$ in the band-index space with $N$, $q$, and $k$ fixed; $({M_{Nk}})_{nm} = [\epsilon^{(\srm)} + 2 h \left( N + {1}/{2} \right) + {\xi_{c}}^2 k^2] \delta_{nm} + \eta (1 - \delta_{nm})$. 
Two normalized eigenvectors of $M_{Nk}$ are given by $\bm{u} = 2^{-1/2} (1, -\text{sgn} \, \eta)^\text{T}$ and $\overline{\bm{u}} = 2^{-1/2} (1, +\text{sgn} \, \eta)^\text{T}$, which respectively belong to the following two eigenvalues: 
\begin{equation}
\left\{
\begin{array}{l}
\displaystyle E_{Nk} = \epsilon + 2 h \left( N + \frac{1}{2} \right) + {\xi_c}^2 k^2 \vspace{5pt} \\
\displaystyle \Ebar_{Nk} = \epsbar + 2 h \left( N + \frac{1}{2} \right) + {\xi_c}^2 k^2, 
\end{array}
\right.
\label{eq:EEbar}
\end{equation}
where new reduced temperatures ($\epsilon = \epsilon^{(\srm)} - |\eta|$ and $\epsbar = \epsilon^{(\srm)} + |\eta|$) are introduced. 
By expanding the fluctuating amplitudes as 
\begin{equation}
\left(
\begin{array}{c}
a_{1Nqk} \\
a_{2Nqk}
\end{array}
\right)
= \bm{u} \, b_{Nqk} + \overline{\bm{u}} \, \bbar_{Nqk},
\label{eq:a1a2}
\end{equation}
therefore, we finally obtain the diagonalized form of $\Fcal_\text{2nd}$: 
\begin{equation}
\Fcal_\text{2nd} 
= \sum_{Nqk} \alpha \left( E_{Nk} |b_{Nqk}|^2 + \Ebar_{Nk} |\bbar_{Nqk}|^2 \right). 
\end{equation}
It follows from Eq. (\ref{eq:EEbar}) that $E_{Nk} < \Ebar_{Nk}$, which means that LEM is described by $b_{Nqk}$, while HEM is done by $\bbar_{Nqk}$. 

Equation (\ref{eq:EEbar}) tells us some characters of the considered system. 
First, the bare critical temperatures of the LEM ($\Tc$) and HEM ($\Tcbar$) are determined as $\epsilon = 0$ and $\epsbar = 0$, respectively. In other words, 
\begin{equation}
\left\{
\begin{array}{l}
\displaystyle \Tc = \e^{|\eta|} \, \Tc^{(\srm)} \\
\displaystyle \Tcbar = \e^{-|\eta|} \, \Tc^{(\srm)}, 
\end{array}
\right. 
\end{equation}
so that $\epsilon = \ln (T / \Tc)$ and $\epsbar = \ln (T / \Tcbar)$. 
Second, Eq. (\ref{eq:EEbar}) indicates that the GL coherence lengths are not affected by $\eta$.\cite{symmetriccase} 

Generally, the reduced temperature of the LEM is less than that of the HEM ($\epsilon < \overline{\epsilon}$). 
Therefore, except in the exceptional case with an extremely small $\eta$, the HEM in zero field behaves as an additional noncritical mode even in the critical region for the LEM (e.g., $\overline{\epsilon} \gtrsim 0.1$ when $h = 0$). Then, one needs a method of introducing an energy cutoff which works for not only the LEM but also the additional HEM. 
Among possible ways of introducing a proper cutoff for the short-wavelength-fluctuation modes, we choose the method consistent with that in the single-band case. That is, we apply the energy-cutoff condition in Ref. \onlinecite{Carballeira} to the HEM as well as LEM [cf. Eq. (\ref{eq:cutoffSB})]: 
\begin{equation}
\left\{
\begin{array}{l}
\displaystyle N_\text{cut}  = \frac{c - \epsilon}{2h} - 1, \ 
 k_\text{cut} = \frac{\sqrt{c-\epsilon}}{\xi_c} \vspace{5pt} \\
\displaystyle \overline{N}_\text{cut}  = \frac{c - \overline{\epsilon}}{2h} - 1, \ 
 \overline{k}_\text{cut} = \frac{\sqrt{c-\overline{\epsilon}}}{\xi_c}. 
\end{array}
\right.
\label{eq:cutoff}
\end{equation}
Here $(N_\text{cut}, k_\text{cut})$ for the LEM and $(\Nbar_\text{cut}, \kbar_\text{cut})$ for the HEM are the upper limits of the LLs and wavenumber. 
Note that $c = \mathcal{O}(1)$ is the only dimensionless cutoff parameter we have introduced. 

The mode-coupling effect stemming from the fourth-order term $\Fcal_\text{4th}$ is similar to that in the single-band case. In the present two-band case, 
the interaction between the LEM and HEM also exists in addition to that within each of LEM and HEM. 
It is preferable to treat both LEM and HEM on equal footing, especially when we investigate fluctuation effects in the vortex-liquid region below the renormalized depairing field $H_\text{c2}$ under a finite magnetic field. In fact, if we treat 
HEM in the Gaussian approximation in spite of renormalizing the LEM fluctuation, the resulting magnetization under finite fields would show an unphysical divergence when $\epsbar + h = 0$, i.e., $T = \e^{-h} \Tcbar \approx (1 - h) \Tcbar$, due to neglecting the renormalization of HEM fluctuation. 
Fortunately, in the present symmetric two-band case, we can accomplish the fluctuation renormalization of both LEM and HEM by straightforwardly extending the variational method in the HF approximation for the single-band case as follows. 

First, we prepare two variational parameters $\mu$ and $\mubar$, and divide the GL functional into two parts: $\Fcal = \Fcal_0 + \Fcal_1$, where
\begin{eqnarray}
\Fcal_0 &=& \sum_{N,q,k} \alpha \left[ \mu + 2h \left( N + \frac{1}{2} \right) + {\xi_c}^2 k^2 \right] |b_{Nqk}|^2 \nonumber \\
&& + \sum_{{\overline{N}},{\overline{q}},{\overline{k}}} \alpha \left[ \overline{\mu} + 2h \left( {\overline{N}} + \frac{1}{2} \right) + {\xi_c}^2 {\overline{k}}^2 \right] |\overline{b}_{{\overline{N}}{\overline{q}}{\overline{k}}}|^2. 
\hspace{20pt}
\label{eq:F0}
\end{eqnarray}
Here the LL index and wavevector of the HEM are written as $\Nbar$, $\qbar$, and $\kbar$ so that the upper limits of $(N,k)$ and $(\Nbar, \kbar)$ are respectively $(N_\text{cut}, k_\text{cut})$ and $(\Nbar_\text{cut}, \kbar_\text{cut})$, given in Eq. (\ref{eq:cutoff}). 
We now define a dimensionless temperature $t$ by using the bare critical temperature $\Tc$ for LEM as $t = T / \Tc$. 
Then a trial free-energy density $f_\text{tri} (t, h; \mu, \mubar)$ that is always higher than the exact free-energy density $f_\text{ex} (t, h)$ is defined as 
\begin{eqnarray}
f_\text{ex} 
&=& -(T / V) \ln \text{Tr}_{b,\bbar} \, \e^{-\Fcal / T} \nonumber \\
&=& -(T / V) \ln \text{Tr}_{b,\bbar} \, \e^{-\Fcal_0 / T}
 -(T / V) \ln \langle \e^{-\Fcal_1 / T} \rangle_0 \nonumber \\ &\leq& f_0 + \frac{\langle \Fcal_1 \rangle_0}{V} \equiv f_\text{tri}, 
\end{eqnarray}
where $\langle \cdots \rangle_0 = \text{Tr}_{b, \bbar} \, (\cdots) \, \e^{-\Fcal_0 / T} / \, \text{Tr}_{b, \bbar} \, \e^{-\Fcal_0 / T}$ is the canonical average with respect to $\Fcal_0$. 
Note also that $f_0 = -(T / V) \ln \text{Tr}_{b,\bbar} \, \e^{-\Fcal_0 / T}$ is the Gaussian contribution defined through Eq. (\ref{eq:F0}) for the total free-energy density, and that $\langle \Fcal_1 \rangle_0 / V$ is the residual contribution. 
The former [$f_0 (t, h; \mu, \mubar)$] is rewritten in the following form as shown in Appendix \ref{app:expTB}: 
\begin{equation}
f_0 
= \frac{T_{\crm 0}}{2 \pi^2 {\xi_{ab}}^2 \xi_c} \, t h \,
[I_1 (t, h; \epsilon; \mu) + I_1 (t, h; \epsbar; \mubar)],
\label{eq:f0}
\end{equation}
where $I_1 (t, h; \epsilon; \mu)$ is given in Eq. (\ref{eq:I1SB}). 
The latter ($\langle \Fcal_1 \rangle_0 / V$) can be written by using $f_0$ as described in Appendix \ref{app:fresTB}. 
As a consequence, we obtain 
\begin{eqnarray}
f_\text{tri} 
&=& f_0
-(\mu - \epsilon) \frac{\del f_0}{\del \mu}
-(\overline{\mu} - \overline{\epsilon}) \frac{\del f_0}{\del \overline{\mu}} \nonumber \\
&& + \frac{\beta}{2 \alpha^2} \left( \frac{\del f_0}{\del \mu} + \frac{\del f_0}{\del \overline{\mu}} \right)^2. 
\label{eq:ftri}
\end{eqnarray}
We can get the optimal free-energy density $f_\text{opt} (t, h)$ by minimizing $f_\text{tri} (t, h; \mu, \mubar)$ with respect to $\mu$ and $\mubar$. 
The optimizing equations $\del f_\text{tri} / \del \mu = \del f_\text{tri} / \del \mubar = 0$ lead to the equations for $\mu$ and $\mubar$: 
\begin{equation}
\left\{
\begin{array}{l}
\displaystyle \mu = \epsilon + \frac{\beta}{\alpha^2} \left(  \frac{\del f_0}{\del \mu} + \frac{\del f_0}{\del \overline{\mu}} \right) \vspace{5pt} \\
\displaystyle \overline{\mu} = \overline{\epsilon} + \frac{\beta}{\alpha^2} \left(  \frac{\del f_0}{\del \mu} + \frac{\del f_0}{\del \overline{\mu}} \right). 
\end{array}
\right.
\label{eq:OPeq}
\end{equation}
We refer to $\mu$ and $\mubar$ that satisfy these equations as $\mu (t, h)$ and $\mubar (t, h)$ hereafter. 

The magnetization is calculated from $M_\text{dia} (t, h) = -(2 \pi {\xi_{ab}}^2 / \phi_0) \del f_\text{opt} (t, h) / \del h$. 
This formula is simplified as follows: 
\begin{eqnarray}
&& M_\text{dia} (t, h) \nonumber \\
&& = - \frac{2 \pi {\xi_{ab}}^2}{\phi_0} 
\frac{\del f_\text{tri} (t, h; \mu, \mubar)}{\del h} \bigg|_{\mu = \mu (t, h), \mubar = \mubar (t, h)} \nonumber \\
&& = - \frac{2 \pi {\xi_{ab}}^2}{\phi_0} 
\frac{\del f_0 (t, h; \mu, \mubar)}{\del h} \bigg|_{\mu = \mu (t, h), \mubar = \mubar (t, h)}. 
\label{eq:M}
\end{eqnarray}
In the first line, we use the fact that $f_\text{opt} (t, h) = f_\text{tri} [t, h; \mu (t, h), \mubar (t, h)]$ is minimized with respect to $\mu$ and $\mubar$. 
In the second line, Eqs. (\ref{eq:ftri}) and (\ref{eq:OPeq}) are used. 
By combining Eqs. (\ref{eq:OPeq}) and (\ref{eq:M}) with the explicit form of $f_0$, we obtain the magnetization as a function of dimensionless temperature $t$ and magnetic field $h$. 
The detailed expression of $M_\text{dia} (t, h)$ is summarized in Appendix \ref{app:expTB}. 
Note that, just as in the single-band case, we can formally push the cutoff $c$ to infinity in the optimizing equations (\ref{eq:OPeq}) and magnetization formula (\ref{eq:M}) by favoring not $\Tc$ (and $\Tcbar$) but the renormalized zero-field critical temperature $T_\text{c} (0)$ as the characteristic temperature scale (see Appendix \ref{app:cindepTB}). 

%%%%%%%%%%%%%%%%%

\section{Results and Discussion}
\label{sec:results}

In this section, the calculated magnetization ($M_\text{dia}$) in a two-band system is presented and compared with that in a single-band system. 
We find that the fluctuation-induced diamagnetism can be enhanced further in the two-band case than that in the single-band case, due to the presence of fluctuations of HEM (see Sec. \ref{sec:methodTB} about the HEM). 

First, some parameters used in evaluating the magnetization are explained, and we will perform comparison between the single-band and two-band cases (Sec. \ref{sec:parameters}). 
Next, we show the magnetization curve as a function of temperature (Sec. \ref{sec:Tdep}) and magnetic field (Sec. \ref{sec:Hdep}). 
Then, it is suggested that the scaling law of magnetization, which is known to appear in a single-band system where fluctuations are strong, can break down in a two-band system (Sec. \ref{sec:scaling}). 
Lastly, after mentioning the interband-coupling dependence of magnetization in the two-band case (Sec. \ref{sec:etadep}), we compare our results in the two-band case with the experimental magnetization observed on FeSe (Sec. \ref{sec:discussion}). 

\subsection{Important Parameters}
\label{sec:parameters}

Material parameters need to be fixed in calculating the magnetization in a single-band system. 
The parameter $c$ in Eq. (\ref{eq:cutoffSB}) means the high-energy cutoff, restricting the short-wavelength fluctuations. 
We ascertain that the change of $c$ within $\mathcal{O}(1)$ does not qualitatively affect the following results. 
The fluctuation strength $\gamma^{(\srm)} \propto \beta$ is defined by Eq. (\ref{eq:gammaSB}), 
\begin{equation}
\gamma^{(\srm)} = \frac{1}{2 \pi^2} \frac{\beta}{\alpha^2} 
\frac{\Tc^{(\srm)}}{{\xi_{ab}}^2 \xi_c}
= \frac{1}{2 \pi^2} \frac{1}{\Delta C^{(\srm)} \, {\xi_{ab}}^2 \xi_c}, 
\nonumber
\end{equation}
which is nothing but the square-root of the Ginzburg number in 3D case (see Appendix \ref{app:exp}). 
Here $\Delta C^{(\srm)}$ is the zero-field specific-heat jump at a mean-field level. 

We move to the case of a two-band system. 
In this case, we fix the interband coupling $\eta$ in Eq. (\ref{eq:Fcal}) as well as the cutoff $c$ in Eq. (\ref{eq:cutoff}) and the fluctuation interaction $\gamma$ (see Appendix \ref{app:expTB}) corresponding to $\gamma^{\rm SB}$ mentioned above and defined by Eq. (\ref{eq:gamma}) is 
\begin{eqnarray}
\gamma = \frac{1}{4 \pi^2} \frac{\beta}{\alpha^2} 
\frac{\Tc}{{\xi_{ab}}^2 \xi_c} 
= \frac{1}{2 \pi^2} \frac{1}{\Delta C^\text{LEM} \, {\xi_{ab}}^2 \xi_c}. 
\nonumber
\end{eqnarray}
Here $\Delta C^\text{LEM}$ is the zero-field specific-heat jump at a mean-field level in the two-band case with a finite interband coupling ($\eta \neq 0$). 

In comparing the magnetization in the single-band and two-band case, we assume that the cutoff ($c$), the bare critical temperature ($\Tc^{(\srm)}$ or $\Tc$), the GL coherence lengths (both $\xi_{ab}$ and $\xi_c$), and the specific-heat jump ($\Delta C^{(\srm)}$ or $\Delta C^\text{LEM}$) take the same values in both cases. 
Referring to the recent experimental data of FeSe,\cite{Terashima, Lin} which indicate values of the GL coherence length and specific-heat jump, we obtain the fluctuation interaction strength as $\gamma^\text{exp} \approx 0.001-0.01$. 
In the following, the parameters are thus set as $c=0.6$, $\gamma^{(\srm)} = \gamma = 0.001$, and $\eta = 0.05$ unless other choice is explicitly given. Under these values of the fluctuation strength, the reduction of the zero-field transition temperature due to the high-energy fluctuations is quite small. 
Regarding $\eta$, this choice results in the relation that $\Tcbar = \e^{-2|\eta|} \, \Tc \simeq 0.90 \Tc$, where $\Tc$ and $\Tcbar$ are, respectively, the bare critical temperatures of LEM and HEM. 
We comment on the limiting cases on the $\eta$ value in Sec. \ref{sec:etadep}. 

%%%%%%%%%%%%%%%%%

\begin{figure}[tbp]
\begin{center}
\includegraphics[scale=0.65]{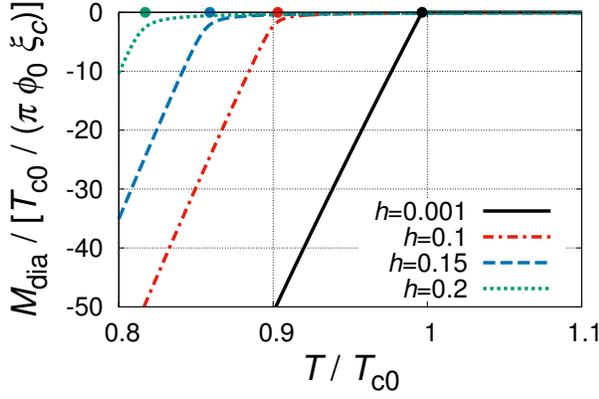}
\caption{(Color online) Temperature ($t = T / \Tc$) dependence of magnetization ($M_\text{dia}$) with varying magnetic fields ($h = 2 \pi {\xi_{ab}}^2 \mu_0 H/\phi_0$) in the two-band case. 
$M_\text{dia}$ is divided by $\Tc / (\pi \phi_0 \xi_c)$ so that it becomes dimensionless. The temperature 
$T_\crm (H)$, which corresponds to $H_\text{c2} (T)$ for each $h$, is also shown with the colored filled circle. 
Parameters are fixed as $c = 0.6$, $\gamma = 0.001$, and $\eta = 0.05$. }
\label{fig:M_Tdep_rough}
\end{center}
\end{figure}

%%%%%%%%%%%%%%%%%

\begin{figure}[tbp]
\begin{center}
\includegraphics[scale=0.65]{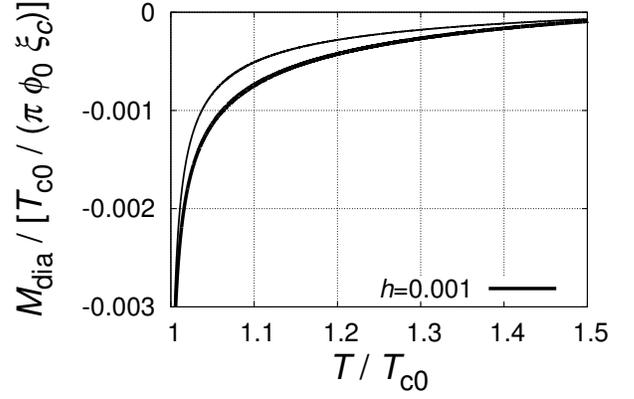}
\caption{Temperature dependence of magnetization at a low field ($h = 0.001$) in the two-band case (with a thick line), compared with the single-band case (with a thin line). 
Parameters ($c$, $\gamma$, and $\eta$) are fixed and symbols are used in the same way as in Fig. \ref{fig:M_Tdep_rough}. }
\label{fig:M_Tdep_lowh}
\end{center}
\end{figure}

%%%%%%%%%%%%%%%%%

\begin{figure}[tbp]
\begin{center}
\includegraphics[scale=0.65]{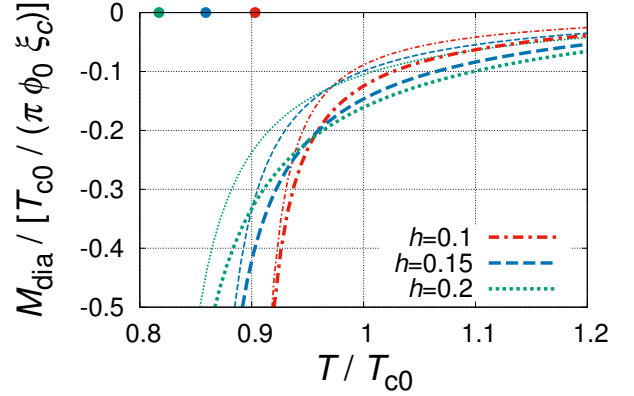}
\caption{(Color online) Temperature dependence of magnetization at high fields ($h = 0.1$, $0.15$, $0.2$) in the two-band case (with thick lines), compared with the single-band case (with thin lines). 
$T_\crm (H)$ for each $h$ is also shown with the colored filled circle. 
Parameters ($c$, $\gamma$, and $\eta$) are fixed and symbols are used in the same way as in Fig. \ref{fig:M_Tdep_rough}. }
\label{fig:M_Tdep_highh}
\end{center}
\end{figure}

%%%%%%%%%%%%%%%%%

\subsection{Temperature Dependence of Magnetization}
\label{sec:Tdep}

Figure \ref{fig:M_Tdep_rough} shows the magnetization expressed as a function of temperature ($t = T / \Tc$) over a broad temperature range in the two-band case, where $h = 2 \pi {\xi_{ab}}^2 \mu_0 H/\phi_0$. 
In this figure, a colored filled circle represents the temperature $T_\crm (H)$ corresponding to $H_\text{c2} (T)$ explained in Sec. \ref{sec:MM} and defined in Appendix \ref{app:TcH} for each magnetization curve shown with the same color. 
See Appedix \ref{app:TcH} regarding the manner of determining the value of $T_\crm (H)$. 
Especially in high fields ($h \gtrsim 0.1$), $M_\text{dia}$ around $T_{\crm} (H)$ is smooth in contrast to the familiar nonanalytic behavior resulting from the mean-field approximation (see, for example, Sec. 5.3. in Ref. \onlinecite{Tinkham}) completely neglecting the fluctuation. As mentioned in Sec. \ref{sec:methodSB}, the mode-coupling effect, i.e., the interaction between fluctuations, produces such smoothing. 
We find that, when is seen over such broad field and temperature ranges as in Fig. \ref{fig:M_Tdep_rough}, the magnetization in the two-band case is hardly distinguishable from that in the single-band case, implying that a two-band character is not reflected in the `rough behavior' of diamagnetism. 

In contrast to the `rough behavior' of the magnetization mentioned in the previous paragraph, the fluctuation contribution to this quantity is affected significantly. 
Figure \ref{fig:M_Tdep_lowh} shows the magnetization curve at a low field ($h = 0.001$). 
It suggests that $|M_\text{dia}|$ is larger in the two-band case, drawn as a thick line, than that in the single-band case (thin line). This result implies that a two-band character seems to appear in the fluctuation-induced diamagnetism. 

In high fields, as well as in low fields, a two-band character emerges in the fluctuation-induced behavior. 
Figure \ref{fig:M_Tdep_highh} shows the magnetization curves in high fields ($h = 0.1$, $0.15$ and $0.2$). 
As a reference, $T_\crm (H)$ for each $h$ is also shown in terms of the colored filled circle as in Fig. \ref{fig:M_Tdep_rough}. 
We see from Fig. \ref{fig:M_Tdep_highh} that the fluctuation-induced diamagnetism in the two-band case (thick lines) is enhanced further in the two-band case than that in the single-band case (thin lines). 
In addition, it is suggested that the magnetization curves approximately cross at a certain temperature (the so-called crossing or intersection point\cite{Rosenstein, Tesanovic, Jiang, PJLin}) with one another not only in the single-band case but also in the two-band case. 
In the two-band case, the crossing point seems to shift to lower temperature compared that with the single-band case. Further, our calculation indicates that the crossing behavior is also visible in 3D case. Previously, the presence of a crossing point of the magnetization curves in high $T_\crm$ cuprates has been argued based on the nearly 2D model and by focusing only on the lowest-LL fluctuation modes.\cite{Tesanovic} However, it has been noted \cite{Oregon} that the use of the lowest-LL approximation is theoretically insufficient for the magnetization data in high $T_c$ cuprates in the tesla range of the applied field. Our numerical results support the validity of the previous argument \cite{Oregon} (see also Appendix D). 

Next, let us comment on the reason why a two-band character appears not in the `rough behavior' (Fig. \ref{fig:M_Tdep_rough}) but in the fluctuation-induced behavior (Figs. \ref{fig:M_Tdep_lowh} and \ref{fig:M_Tdep_highh}). 
First, the `rough behavior' is determined primarily by excitation modes with lower energy. 
In a two-band system, there are two distinct modes, i.e., LEM and HEM (see Sec. \ref{sec:methodTB}). 
The LEM, which has lower energy than that in the HEM, thus produces the `rough behavior'. 
On the other hand, the excitation energy of LEM [$E_{Nk}$ in Eq. (\ref{eq:EEbar})] takes the same form as that in the single-band case\cite{symmetriccase} [Eq. (\ref{eq:F2ndSB})]. 
Therefore, the `rough behavior' of diamagnetism in the two-band case becomes essentially the same as that in the single-band case. 

By contrast, excitation modes with higher energy can contribute to the fluctuation-induced behavior. 
In a two-band system, fluctuations of the HEM thus enhance the total diamagnetism in addition to those of the LEM. 
Since the excitaion energy of the LEM [$E_{Nk}$ in Eq. (\ref{eq:EEbar})] is the same as that of the single-band case\cite{symmetriccase} [Eq. (\ref{eq:F2ndSB})], the fluctuation-induced diamagnetism is enhanced further in the two-band case than that in the single-band case. 

%%%%%%%%%%%%%%%%%

\begin{figure}[tbp]
\begin{center}
\includegraphics[scale=0.65]{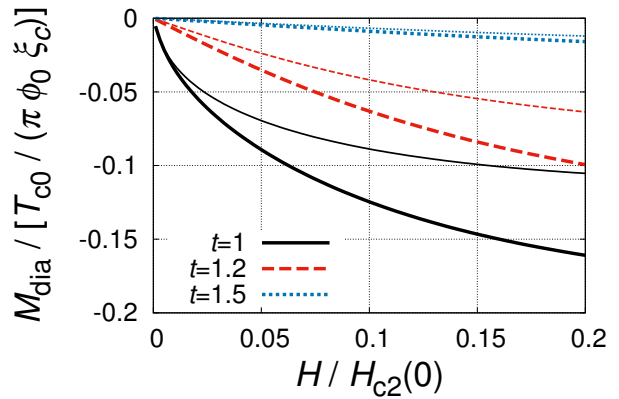}
\caption{(Color online) Magnetic-field dependence of magnetization at different temperatures ($t = T / \Tc = 1$, $1.2$, $1.5$) in the two-band case (thick lines) and in the single-band case (thin lines). 
Parameters ($c$, $\gamma$, and $\eta$) are fixed and other symbols are used in the same way as in Fig. \ref{fig:M_Tdep_rough}. }
\label{fig:M_hdep}
\end{center}
\end{figure}

%%%%%%%%%%%%%%%%%

\subsection{Magnetic-Field Dependence of Magnetization}
\label{sec:Hdep}

To further investigate a two-band character, the magnetic-field dependence of magnetization is shown in Fig. \ref{fig:M_hdep} with varying temperatures as $t = T / \Tc = 1$, $1.2$, and $1.5$. 
As in Figs. \ref{fig:M_Tdep_lowh} and \ref{fig:M_Tdep_highh}, thick and thin lines represent the magnetization curves in the two-band and single-band cases, respectively. 
We see from Fig. \ref{fig:M_hdep} that the difference in magnetization between the two-band and single-band cases becomes larger in higher magnetic fields, especially when $T \gtrsim \Tc$ ($t \gtrsim 1$). 
This feature is explained as follows. 

When $h \rightarrow 0$ and the temperature is above $\Tc$ and outside the critical region, the diamagnetic susceptibility is known to show a divergent behavior as a function of temperature; 
For example, in a single-band three-dimensional system, the magnetization $M_\text{dia}^{(\srm)}$ behaves\cite{Schmid} as
\begin{equation}
\frac{M_\text{dia}^{(\srm)} (t, h)}{h} \simeq - \frac{\Tc^{(\srm)}}{12 \phi_0 \xi_c} \frac{t}{\sqrt{t - 1}}. 
\end{equation}
This divergence is due to the low-energy (long-wavelength) fluctuations. 
In the present two-band case, long-wavelength fluctuations in LEM play a role in producing such divergence, and thus LEM becomes much more significant than HEM when $t \gtrsim 1$ and $h \rightarrow 0$. 
Therefore, when $t \gtrsim 1$ and $h \rightarrow 0$, the fluctuation-induced diamagnetism is dominated by LEM, which behaves in the same manner as the single-band case within our model.\cite{symmetriccase} 
On the other hand, when $t \gtrsim 1$ and $h$ gets higher, fluctuations of HEM tend to more substantially contribute to the diamagnetism. 
As a consequence, diamagnetism in the two-band case becomes more enhanced in higher fields, compared with the single-band case. 

%%%%%%%%%%%%%%%%%

\subsection{Breakdown of Scaling Law}
\label{sec:scaling}

In a single-band system with strong fluctuations, it is known that characteristic scaling laws of the thermodynamic quantities and transport coefficients appear in high fields,\cite{Welp, Ikeda, Tesanovic, Ullah} due to the restriction of the fluctuation modes to the lowest LL. 
As for the fluctuation-induced diamagnetism of our interest in this study, the magnetization $M_\text{dia}^{(\srm)}$ in a single-band system is scaled as\cite{Welp}
\begin{equation}
\frac{M_\text{dia}^{(\srm)} (t, h)}{\left( t h / \sqrt{\gamma^{(\srm)}} \right)^{2/3}} \propto - S \left[ \frac{t - t_\crm (h)}{(\gamma^{(\srm)} \, t h)^{2/3}} \right]. 
\label{eq:scaling}
\end{equation}
Here $S(x)$ is a certain scaling function, and $t_\crm^{(\srm)} (h) = T_\crm^{(\srm)} (H) / \Tc^{(\srm)}$. The region in the field v.s. temperature phase diagram where this lowest-LL scaling behavior should be visible has been discussed in the single-band case in Ref. \onlinecite{RI95} in details.

\begin{figure}[tbp]
\begin{center}
\includegraphics[scale=0.65]{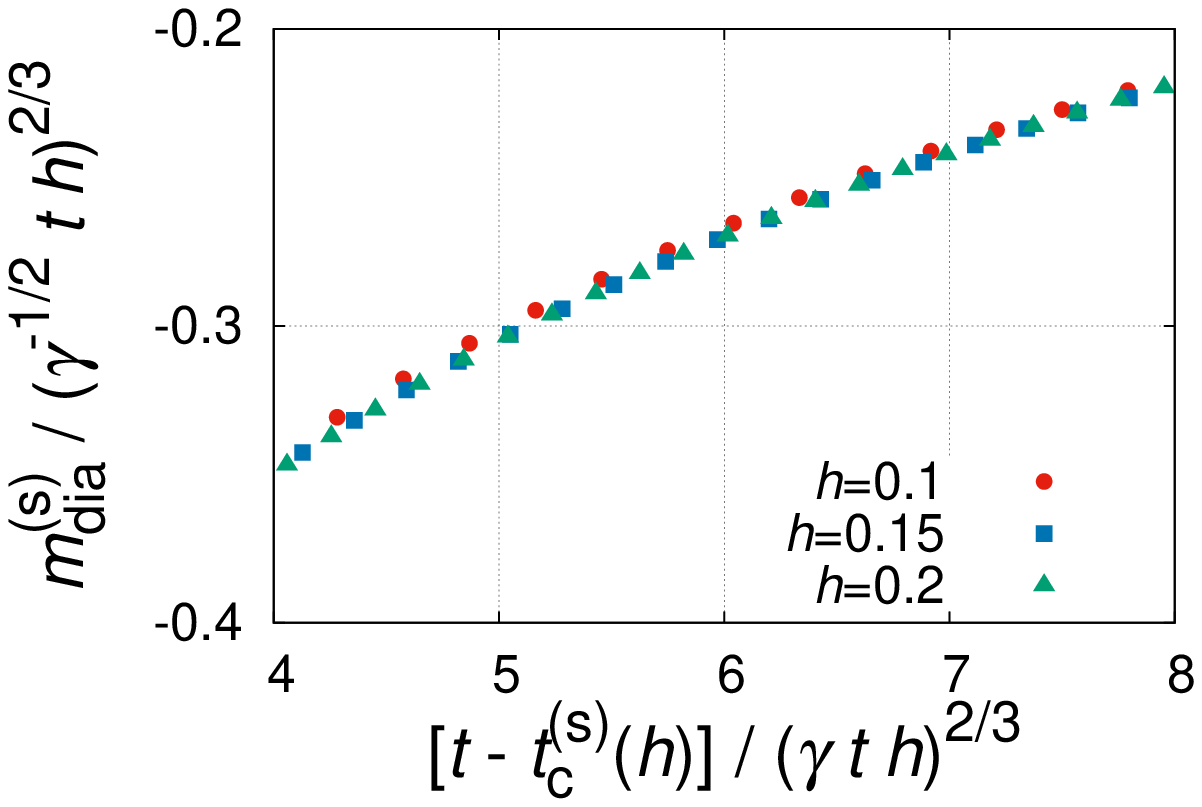}
\caption{(Color online) Scaling plot of magnetization in the single-band case at high fields ($h = 0.1$, $0.15$, $0.2$). 
A dimensionless magnetization $m_\text{dia}^{(\srm)} = M_\text{dia}^{(\srm)} / [\Tc / (\pi \phi_0 \xi_c)]$ is introduced. 
Parameters ($c$ and $\gamma$) are fixed and other symbols are used in the same way as in Fig. \ref{fig:M_Tdep_rough}. }
\label{fig:M_scaling_SB}
\end{center}
\end{figure}

\begin{figure}[tbp]
\begin{center}
\includegraphics[scale=0.65]{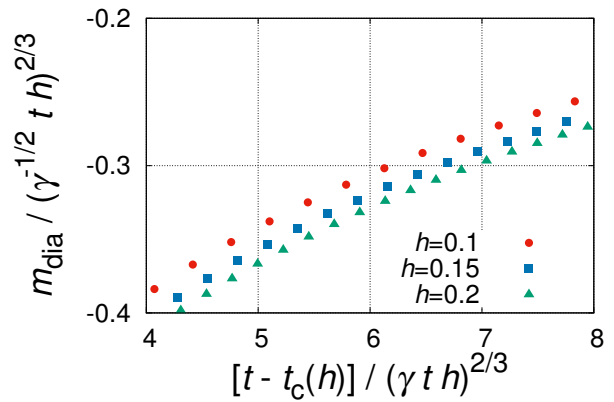}
\caption{(Color online) Scaling plot of magnetization in the two-band case at high fields ($h = 0.1$, $0.15$, $0.2$). 
A dimensionless magnetization $m_\text{dia} = M_\text{dia} / [\Tc / (\pi \phi_0 \xi_c)]$ is introduced. 
Parameters ($c$, $\gamma$, and $\eta$) are fixed and symbols are used in the same way as in Fig. \ref{fig:M_Tdep_rough}. 
}
\label{fig:M_scaling_TB}
\end{center}
\end{figure}

In our two-band case, fluctuations of HEM as well as LEM can contribute to the diamagnetism (see Secs. \ref{sec:Tdep} and \ref{sec:Hdep}). 
Thus the scaling law [Eq. (\ref{eq:scaling})] may be broken by fluctuations of HEM. 
In fact, by comparing Figs. \ref{fig:M_scaling_SB} (a scaling plot for the single-band case) and \ref{fig:M_scaling_TB} (a scaling plot for the two-band case) with each other, we clearly see that a breakdown of the scaling law occurs in the two-band case. 

%%%%%%%%%%%%%%%%%

%%%%%%%%%%%%%%%%%

\subsection{Interband-Coupling Dependence of Magnetization}
\label{sec:etadep}

Here, we briefly explain the interband-coupling ($\eta$) dependence of the magnetization. 
First, when $\eta \gg 1$ with a fixed $\Tc$, fluctuations of HEM are completely suppressed, and the property of a two-band system is determined by LEM. 
Since the present two-band case with no contribution of HEM is equivalent to the single-band case as mentioned in Secs. \ref{sec:Tdep} and \ref{sec:Hdep}, the resulting magnetization in the two-band case coincides with that in the single-band case. 
Because a finite cutoff $c$ was introduced {\it practically} in the present approach, effects of the HEM fluctuations disappear when $\eta$ exceeds 
a certain value. 
Next, the case where $\eta \rightarrow 0$ with a fixed $\Tc$ will be commented on. 
In this case, the two bands are completely separated and behave independently. 
Above the critical region,\cite{eta0lowT} therefore, we obtain twice the magnetization of that in the single-band case. 
Lastly, in the intermediate range of $\eta$ values, such as our choice ($\eta = 0.05$) in the previous sections, we obtain the magnetization not less than one time and not more than twice that in the single-band case. 

%%%%%%%%%%%%%%%%%

\subsection{Discussion}
\label{sec:discussion}

In this section, we discuss the relation between our magnetization results in a two-band system and the large diamagnetism observed recently in FeSe.\cite{Kasahara}

Since, in Ref. \onlinecite{Kasahara}, the measured diamagnetic susceptibility ($\chi_\text{dia}^\text{exp} = M^\text{exp}_\text{dia} / H$) is plotted as a function of temperature, we compare $\chi_\text{dia}^\text{exp}$ with the corresponding quantity calculated in both single-band and two-band cases ($\chi_\text{dia}^{(\srm)} = M_\text{dia}^{(\srm)} / H$ and $\chi_\text{dia} = M_\text{dia} / H$, respectively).\cite{material} 
As a result, the calculated susceptibility in the two-band case is closer to the experimental value than that in the single-band case: $|\chi^\text{exp}_\text{dia}| > |\chi_\text{dia}| > |\chi^{(\srm)}_\text{dia}|$. 
However, $|\chi_\text{dia}^\text{exp}|$ is still two or three times larger than $|\chi_\text{dia}|$. 
This fact indicates that in order to quantitatively understand the anomalously large diamagnetism on FeSe, we should take account of  factors other than the simple two-band structure. 
As a candidate of such factors, it will be interesting to theoretically incorporate the argument\cite{Kasahara,Kasahara2} that FeSe is possibly in the so-called BCS-BEC-crossover regime. 
We also suggest other possibilities such that the interband asymmetry neglected in our model may be important or that an additional small electron pocket\cite{Huynh, Watson} may contribute to SCFs. 

%%%%%%%%%%%%%%%%%

\section{Summary and Conclusions}
\label{sec:summary}

The anomalously large diamagnetism observed recently in FeSe motivates us to study how a two-band character in this material affects the SCF-induced behavior. 
We investigate the SCF-induced diamagnetism in a two-band system on the basis of a phenomenological approach, starting with a symmetric two-band GL functional. 
When we consider SCFs, it is important that a two-band system has two distinct fluctuation modes, i.e., the LEM with lower excitation energy and the HEM with higher excitation energy. 
As long as asymmetry between the two bands is small, the fluctuations of the LEM behave in a similar way to those in a corresponding single-band system; 
On the other hand, the HEM is unique to a two-band system and enhances the SCF-induced diamagnetism. 
This enhancement is summarized in the next paragraph as follows. 

First, we ascertain that the `rough behavior' of magnetization is similar to that in a single-band system. 
This is due to the fact that the `rough behavior' is attributed largely to the LEM. 
Second, both in low and high fields, the SCF-induced diamagnetism is found to be enhanced further in a two-band system than in a single-band system. 
This additional enhancement of SCF specific to a two-band system is caused by the fluctuations of the HEM. 
Furthermore, we find that, as far as the temperature is near the zero-field critical temperature, the diamagnetism stemming from the HEM fluctuations becomes more remarkable in higher fields. 
Third, we also focus on the behavior specific to a high-field region; 
We examine whether or not the crossing point of magnetization curves and the scaling behavior\cite{Ikeda,Ullah,Welp} of the magnetization appear in a two-band system. 
As a result, the crossing point moves to lower temperature as compared with the single-band case, and the scaling behavior seems to break down. 
These features are attributed again to the HEM fluctuations. 
Lastly, we compare our results with the magnetization observed in FeSe. 
The calculated magnetization in the two-band case is closer to the experimental data than that in the single-band case; 
However, quantitative agreement seems to require a certain mechanism other than the simple two-band structure, such as the characteristic feature of a system in the BCS-BEC-crossover regime. 

%%%%%%%%%%%%%%%%%

\appendix

\section{Single-Band Case}
\label{app:SB}

In the following, we describe some details omitted in the main text in the single-band case. 

\subsection{Explicit Form of Some Quantities}
\label{app:exp}

In this appendix, the explicit form of some quantities introduced in the main text is provided: 
The Gaussian contribution of the renormalized fluctuations [Eq. (\ref{eq:f0SB})], the optimizing equation [Eq. (\ref{eq:OPeqSB})], and the magnetization [Eq. (\ref{eq:MSB})]. 

The Gaussian contribution of the renormalized fluctuations to the whole free-energy density [$f_0^{(\srm)} = -(T / V) \ln \text{Tr}_{a} \, \e^{-\Fcal_0^{(\srm)} / T}$] is calculated as follows. 
First, the Gaussian integral leads to
\begin{equation}
f_0^{(\srm)} 
= - \frac{T}{V} \sum_{N,q,k} 
\ln \frac{\pi T}{\alpha \left[ \mu + 2 h (N + 1/2) + {\xi_c}^2 k^2 \right]}. 
\end{equation}
Next, recalling that each LL is ($\mu_0 H L_x L_y / \phi_0$)-fold degenerate and that the dimensionless temperature and magnetic field are respectively defined as $t = T / T_\text{c0}^{(\srm)}$ and $h = 2 \pi {\xi_{ab}}^2 \mu_0 H / \phi_0$, we obtain
\begin{eqnarray}
&& f_0^{(\srm)}
= - \frac{\Tc^{(\srm)}}{4 \pi^2 {\xi_{ab}}^2} \, th 
\nonumber \\
&& \hspace{30pt} \times
\int _{- k_\text{cut}} ^{k_\text{cut}} \drm k \sum_{N=0}^{N_\text{cut}} 
\ln \frac{\pi \Tc^{(\srm)} \, t}{\alpha \left[ \mu + 2 h (N + 1/2) + {\xi_c}^2 k^2 \right]}. 
\nonumber
\end{eqnarray}
Note that here the summation with respect to $k$ is approximated to the integral: $\sum_k \simeq (L_z / 2 \pi) \int \drm k$. 
Lastly, we use the fact that the Gamma function $\Gamma (x)$ satisfies
\begin{equation}
\prod_{N=0}^{N_\text{cut}} (N+X)
= \frac{\Gamma (N_\text{cut} + X + 1)}{\Gamma (X)}
\label{eq:Gamma}
\end{equation}
with the cutoff condition for $N$ and $k$ [Eq. (\ref{eq:cutoffSB})], leading to
\begin{equation}
f^{(\srm)}_0
= \frac{T_{\crm 0}^{(\srm)}}{2 \pi^2 {\xi_{ab}}^2 \xi_c} \, t h \,
I_1 (t, h; \epsilon^{(\srm)}; \mu), 
\nonumber 
\end{equation}
where 
\begin{eqnarray}
I_1 (t, h; \epsilon^{(\srm)}; \mu)
&=& \int_0^{\sqrt{c-\epsilon^{(\srm)}}} \drm \widetilde{k} 
\Bigg[
\frac{c-\epsilon^{(\srm)}}{2h} \ln \left( \frac{2\alpha}{\pi \Tc} \frac{h}{t} \right) \nonumber \\
&& \hspace{48pt} + \ln \Gamma \left( \frac{\widetilde{k}^2 + \mu + c -\epsilon^{(\srm)}}{2h} + \frac{1}{2} \right) \nonumber \\
&& \hspace{48pt} - \ln \Gamma \left( \frac{{\widetilde{k}}^2 + \mu}{2h} + \frac{1}{2} \right)
\Bigg]. 
\label{eq:I1SB}
\end{eqnarray}
Here we introduce a dimensionless wavenumber $\widetilde{k} = \xi_c k$ in the final expression. 

The optimizing equation [Eq. (\ref{eq:OPeqSB})] can be rewritten explicitly as follows. 
First, substituting Eq. (\ref{eq:I1SB}) for $I_1$ in $f_0^{(\srm)}$ [Eq. (\ref{eq:f0SB})] and differentiating $f_0^{(\srm)}$ with respect to $\mu$, we get
\begin{eqnarray}
\frac{\del f_0^{(\srm)}}{\del \mu} 
&=& \frac{T_{\crm 0}^{(\srm)}}{2 \pi^2 {\xi_{ab}}^2 \xi_c} \, t h \,
\frac{\del I_1 (t, h; \epsilon^{(\srm)}; \mu)}{\del \mu} \nonumber \\
&=& \frac{T_{\crm 0}^{(\srm)}}{4 \pi^2 {\xi_{ab}}^2 \xi_c} \, t \nonumber \\ 
&& \times \int_0^{\sqrt{c - \epsilon^{(\srm)}}} \drm \widetilde{k} 
\Bigg[
\psi \left( \frac{\ktil^2 + \mu + c - \epsilon^{(\srm)}}{2 h} + \frac{1}{2} \right) \nonumber \\
&& \hspace{60pt} - \psi \left( \frac{\ktil^2 + \mu}{2 h} + \frac{1}{2} \right)
\Bigg]. 
\label{eq:df0dmu}
\end{eqnarray}
Note that we introduce the so-called digamma function defined as $\psi (x) = \drm \ln \Gamma (x) / \drm x$. 
Then, using Eqs. (\ref{eq:OPeqSB}) and (\ref{eq:df0dmu}), we arrive at the final expression: 
\begin{equation}
\mu = \epsilon^{(\srm)} + \gamma^{(\srm)} \, t \, I_2 (t, h; \epsilon^{(\srm)}; \mu), 
\label{eq:OPeqSBexp}
\end{equation}
where
\begin{eqnarray}
I_2 (t, h; \epsilon^{(\srm)}; \mu)
&=& \int_0^{\sqrt{c - \epsilon^{(\srm)}}} \drm \widetilde{k} 
\Bigg[
\psi \left( \frac{\ktil^2 + \mu + c - \epsilon^{(\srm)}}{2 h} + \frac{1}{2} \right) \nonumber \\
&& \hspace{55pt} - \psi \left( \frac{\ktil^2 + \mu}{2 h} + \frac{1}{2} \right)
\Bigg]. 
\label{eq:I2SB}
\end{eqnarray}
Here $\gamma^{(\srm)}$ is an important parameter that represents the fluctuation strength, defined by
\begin{equation}
\gamma^{(\srm)} = \frac{1}{2 \pi^2} \frac{\beta}{\alpha^2} 
\frac{\Tc^{(\srm)}}{{\xi_{ab}}^2 \xi_c} 
= \frac{1}{2 \pi^2} \frac{1}{\Delta C^{(\srm)} \, {\xi_{ab}}^2 \xi_c}, 
\label{eq:gammaSB}
\end{equation}
where $\Delta C^{(\srm)} = \alpha^2 / (\beta \Tc^{(\srm)})$ is the zero-field specific-heat jump at a mean-field level. 

In this paragraph, the explicit form of the magnetization [Eq. (\ref{eq:MSB})] will be given. 
By combining Eqs. (\ref{eq:f0SB}), (\ref{eq:MSB}), and (\ref{eq:I1SB}) with one another, we obtain 
\begin{eqnarray}
M_\text{dia}^{(\srm)} (t, h)
&=& - \frac{2 \pi {\xi_{ab}}^2}{\phi_0} 
\frac{\Tc^{(\srm)}}{2 \pi^2 {\xi_{ab}}^2 \xi_c} \, t \nonumber \\ 
&& \hspace{40pt} \times \frac{\del [h \, I_1 (t, h; \epsilon^{(\srm)}; \mu)]}{\del h} \Bigg|_{\mu = \mu (t, h)} 
\nonumber \\
&=& - \frac{\Tc^{(\srm)}}{\pi \phi_0 \xi_c} \, t \, I_3 (t, h; \epsilon^{(\srm)}; \mu) \big|_{\mu = \mu (t, h)}, 
\label{eq:MdiaSBexp}
\end{eqnarray}
where
\begin{eqnarray}
&& I_3 (t, h; \epsilon^{(\srm)}; \mu) \nonumber \\
&& = 
\int_0^{\sqrt{c-\epsilon^{(\srm)}}} \drm \ktil \nonumber \\
&& \hspace{25pt} \times
\bigg[ 
\frac{c-\epsilon^{(\srm)}}{2h} + \ln \Gamma \left( \frac{\ktil^2 + \mu + c - \epsilon^{(\srm)}}{2h} + \frac{1}{2} \right) \nonumber \\
&& \hspace{40pt} 
- \frac{\ktil^2 + \mu + c - \epsilon^{(\srm)}}{2h} \psi \left( \frac{\ktil^2 + \mu + c - \epsilon^{(\srm)}}{2h} + \frac{1}{2} \right) \nonumber \\
&& \hspace{40pt} - \ln \Gamma \left( \frac{\ktil^2 + \mu}{2h} + \frac{1}{2} \right)
\nonumber \\
&& \hspace{40pt} + \frac{\ktil^2 + \mu}{2h} \psi \left( \frac{\ktil^2 + \mu}{2h} + \frac{1}{2} \right)  \bigg]. 
\label{eq:I3SB}
\end{eqnarray}

\subsection{Cutoff-Independent Optimizing Equation and Magnetization Formula}
\label{app:cindepSB}

In the following, we show the derivation of the asymptotic form of the optimizing equation [Eq. (\ref{eq:OPeqSBexp})] and magnetization formula [Eq. (\ref{eq:MdiaSBexp})] in $c \rightarrow \infty$ limit. 
Following Ref. \onlinecite{Jiang}, we replace $\epsilon^\text{(s)}$ by $T / \Tc^\text{(s)} - 1$ in this Appendix. 

In order to obtain the zero-field critical temperature $T_\text{c}^\text{(s)} (0)$ as a function of the cutoff $c$, we take the limit $h \rightarrow 0$ and $\mu = 0$ in Eq. (\ref{eq:OPeqSBexp}).
Then, one finds 
\begin{equation}
\frac{\TcSB (0)}{\TczSB} - 1 = - \gamtil^\text{(s:R)} \sqrt{c} - \frac{\left( \gamtil^\text{(s:R)} \right)^2}{2} + \mathcal{O} \left(  \frac{1}{\sqrt{c}} \right), 
\label{eq:ZFTcSB}
\end{equation}
where 
\begin{eqnarray}
\gamtil^\text{(s:R)} &=& \left( \frac{\pi}{2} + \ln 2 \right) \gamma^\text{(s:R)} \no \\
&=& \left( \frac{\pi}{2} + \ln 2 \right) \frac{1}{2 \pi^2} \frac{\beta}{\alpha^2} \frac{\TcSB (0)}{{\xi_{ab}}^2 \xi_c} 
\label{eq:gamtilSB}
\end{eqnarray}
represents the fluctuation strength just at the renormalized critical temperature $\TcSB (0)$. 

Using the asymptotic form of the digamma function [$\psi (x) \rightarrow \ln x - (2 x)^{-1} + \mathcal{O} (x^{-2})$ $(x \rightarrow \infty)$] in (\ref{eq:I2SB}), we can arrive at the asymptotic form of $I_2$ as $c \rightarrow \infty$: 
\begin{eqnarray}
I_2 (t, h; \epsilon^\text{(s)}; \mu) &=& - \hspace{-3pt} \int _0 ^\infty \hspace{-3pt} \drm \ktil \ \overline{\psi} \bigg( \frac{{\ktil}^2 + \mu}{2 h} + \frac{1}{2} \bigg) \no \\
&& + \left( \frac{\pi}{2} + \ln 2 \right) \left( \sqrt{c} + \frac{\gamtil^\text{(s:R)} \, t^\text{(R)}}{2} \right) \no \\
&& - \pi \sqrt{\mu + h} + \mathcal{O} \left( \frac{\ln c}{\sqrt{c}} \right), 
\label{eq:I2asymptSB}
\end{eqnarray}
where
\begin{equation}
\overline{\psi} (x) = \psi (x) - \ln x.
\label{eq:psibar}
\end{equation}
Here, we define a dimensionless temperature $t^\text{(R)} = T / \TcSB (0)$. 

Combining Eqs (\ref{eq:OPeqSBexp}), (\ref{eq:ZFTcSB}), and (\ref{eq:I2asymptSB}) altogether, one can finally obtain the following formula: 
\begin{eqnarray}
\mu &=& t^\text{(R)} - 1 \no \\
&& + \gamma^\text{(s:R)} \, t^\text{(R)} \bigg[ \hspace{-3pt} - \hspace{-3pt} \int _0 ^\infty \hspace{-3pt} \drm \ktil \ \overline{\psi} \bigg( \frac{{\ktil}^2 + \mu}{2 h} + \frac{1}{2} \bigg) \hspace{-1pt} - \hspace{-1pt} \pi \sqrt{\mu + h} \hspace{15pt} \no \\
&& \hspace{53pt} + \frac{1}{2} \left( \frac{\pi}{2} + \ln 2 \right) \gamtil^\text{(s:R)} \left( t^\text{(R)} - 1 \right) \bigg]. 
\label{eq:OPeqasymptSB}
\end{eqnarray}
Equation (\ref{eq:OPeqasymptSB}) is entirely independent of the cutoff parameter $c$ and represented in terms of the renormalized zero-field critical temperature $\TcSB (0)$ instead of the bare critical temperature $\TczSB$. 

As for the behavior of the magnetization formula [Eq. (\ref{eq:MdiaSBexp})] in $c \rightarrow \infty$ limit, the asymptotic form of the gamma function [$\ln \Gamma (x) \rightarrow x \ln x - x - (1/2) \ln x + (1 /2) \ln (2 \pi) + \mathcal{O} (x^{-1})$] will be used in addition to that of the digamma function mentioned above, and then we obtain
\begin{equation}
M_\text{dia}^\text{(s)} = - \frac{\TcSB (0)}{\pi \phi_0 \xi_c} \, t^\text{(R)} \int _0 ^\infty \drm \ktil \, \Upsilon \bigg( \frac{\ktil^2 + \mu}{2 h} + \frac{1}{2} \bigg), 
\label{eq:MdiaasymptSB}
\end{equation}
where
\begin{equation}
\Upsilon (x) = - \ln \Gamma (x) + \left( x - \frac{1}{2} \right) \psi (x) - x + \frac{1}{2} \left[ 1 + \ln (2 \pi) \right]. 
\label{eq:Upsilon}
\end{equation}
Equation (\ref{eq:MdiaasymptSB}), as well as Eq. (\ref{eq:OPeqasymptSB}), is independent of the cutoff $c$ and represented in terms of $\TcSB (0)$ instead of $\TczSB$. 

We only have to neglect the fluctuation interaction [$\gamma^\text{(s:R)} = 0$ in Eq. (\ref{eq:OPeqasymptSB})] or replace $\mu$ with $t^\text{(R)} - 1$ in Eq. (\ref{eq:MdiaasymptSB}) in order to reproduce the Gaussian magnetization formula\cite{Carballeira} equivalent to that obtained in Ref. \onlinecite{Prange}. 
Provided $t^\text{(R)} - 1 \ll h$, in particular, we will arrive at the familiar result: 
\begin{equation}
M_\text{dia}^\text{(s)} = - C \, \frac{T}{\pi \phi_0 \xi_c} \sqrt{h} \propto -\sqrt{h}, 
\end{equation}
where $C \simeq 0.406$. 

\subsection{Trial Free-Energy Density}
\label{app:fres}

In this appendix, we will show how Eq. (\ref{eq:ftriSB}) is derived. 
We recall that $f_\text{tri}^{(\srm)} = f_0^{(\srm)} + \langle \Fcal_1^{(\srm)} \rangle_0 / V$, and thus it is sufficient to write down how we treat $\langle \Fcal_1^{(\srm)} \rangle_0$. 
First, $\Fcal_1^{(\srm)}$ is defined as 
\begin{eqnarray}
\Fcal_1^{(\srm)} &=& \Fcal^{(\srm)} - \Fcal_0^{(\srm)} \nonumber \\
&=& - \alpha (\mu - \epsilon^{(\srm)}) \int_V \drm^3 \rbm |\psi (\rbm)|^2 \nonumber \\
&& + \frac{\beta}{2} \int_V \drm^3 \rbm |\psi (\rbm)|^4. 
\nonumber
\end{eqnarray}
With Eq. (\ref{eq:OPexpSB}), then, we expand $\psi (\rbm)$ as a linear combination of $a_{Nqk}$. 
The property of the Gaussian integral, such as $\langle a_{N q k}^* a_{N q k}^* a_{N q k} a_{N q k} \rangle_0 = 2 ( \langle |a_{N q k}|^2 \rangle_0 )^2$, leads to
\begin{eqnarray}
\langle \Fcal_1^{(\srm)} \rangle_0 
&=& - \alpha (\mu - \epsilon^{(\srm)}) \sum_{N, q, k} \langle |a_{N q k}|^2 \rangle_0 \nonumber \\
&& + \beta \frac{1}{V} 
\Bigg( 
\sum_{N, q, k} \langle |a_{N q k}|^2 \rangle_0
\Bigg)^2. 
\label{eq:F1avSB}
\end{eqnarray}
From $f_0^{(\srm)} = - (T / V) \ln \text{Tr}_{a} \, \e^{- \Fcal^{(\srm)}_0 / T}$, on the other hand, we see that 
\begin{equation}
\frac{\del f^{(\srm)}_0}{\del \mu} 
= \frac{\alpha}{V} \sum_{N, q, k} \langle |a_{N q k}|^2 \rangle_0. 
\label{eq:f0diffSB}
\end{equation}
Combining Eqs. (\ref{eq:F1avSB}) with (\ref{eq:f0diffSB}), we obtain
\begin{equation}
\frac{\langle \Fcal_1^{(\srm)} \rangle_0}{V}
= -(\mu - \epsilon^{(\srm)}) \frac{\del f^{(\srm)}_0}{\del \mu}
+ \frac{\beta}{\alpha^2} \left( \frac{\del f^{(\srm)}_0}{\del \mu} \right)^2. 
\nonumber
\end{equation}
From $f_\text{tri}^{(\srm)} = f_0^{(\srm)} + \langle \Fcal_1^{(\srm)} \rangle_0 / V$, therefore, we reach the expected equation [Eq. (\ref{eq:ftriSB})]: 
\begin{equation}
f_\text{tri}^{(\srm)} = f_0^{(\srm)} - (\mu - \epsilon^{(\srm)}) \frac{\del f^{(\srm)}_0}{\del \mu}
+ \frac{\beta}{\alpha^2} \left( \frac{\del f^{(\srm)}_0}{\del \mu} \right)^2. 
\nonumber
\end{equation}

%%%%%%%%%%%%%%%%%

\section{Two-Band Case}
\label{app:TB}

In the following, we describe some details omitted in the main text in the two-band case. 
Note that basic ideas are the same as in the single-band case (Appendix \ref{app:SB}). 

\subsection{Explicit Form of Some Quantities}
\label{app:expTB}

In this appendix, the explicit forms of the Gaussian contribution of the renormalized fluctuations [Eq. (\ref{eq:f0})], the optimizing equations [Eq. (\ref{eq:OPeq})], and the magnetization [Eq. (\ref{eq:M})] are provided. 

The Gaussian contribution of the renormalized fluctuations to the whole free-energy density [$f_0 = -(T / V) \ln \text{Tr}_{b, \bbar} \, \e^{-\Fcal_0 / T}$] is calculated as follows. 
First, the Gaussian integral leads to
\begin{eqnarray}
f_0 
&=& - \frac{T}{V} \sum_{N,q,k} 
\ln \frac{\pi T}{\alpha \left[ \mu + 2 h (N + 1/2) + {\xi_c}^2 k^2 \right]}  \nonumber \\
&& - \frac{T}{V} \sum_{\Nbar,\qbar,\kbar} 
\ln \frac{\pi T}{\alpha \left[ \mubar + 2 h (\Nbar + 1/2) + {\xi_c}^2 \kbar^2 \right]}. 
\end{eqnarray}
Next, recalling that each LL is ($\mu_0 H L_x L_y / \phi_0$)-fold degenerate and that the dimensionless temperature and magnetic field are respectively defined as $t = T / T_\text{c0}$, we obtain
\begin{eqnarray}
&& f_0
= - \frac{\Tc}{4 \pi^2 {\xi_{ab}}^2} \, th 
\nonumber \\
&& \hspace{20pt} \times \Bigg\{ 
\int _{- k_\text{cut}} ^{k_\text{cut}} \drm k \sum_{N=0}^{N_\text{cut}} 
\ln \frac{\pi \Tc \, t}{\alpha \left[ \mu + 2 h (N + 1/2) + {\xi_c}^2 k^2 \right]} 
\nonumber \\
&& \hspace{30pt} + 
\int _{- \kbar_\text{cut}} ^{\kbar_\text{cut}} \drm \kbar \sum_{\Nbar=0}^{\Nbar_\text{cut}} 
\ln \frac{\pi \Tc \, t}{\alpha \left[ \mubar + 2 h (\Nbar + 1/2) + {\xi_c}^2 \kbar^2 \right]} \Bigg\}, 
\nonumber
\end{eqnarray}
where $\Tc$ is the bare critical temperature of the LEM, and $h = 2 \pi {\xi_{ab}}^2 \mu_0 H / \phi_0$. 
Note that here the summation with respect to $k$ is approximated to the integral: $\sum_k \simeq (L_z / 2 \pi) \int \drm k$. 
Lastly, we use Eq. (\ref{eq:Gamma}) with the cutoff condition for $(N, k)$ and $(\Nbar, \kbar)$ [see Eq. (\ref{eq:cutoff})]. 
Finally, we obtain
\begin{equation}
f_0
= \frac{T_{\crm 0}}{2 \pi^2 {\xi_{ab}}^2 \xi_c} \, t h 
\left[ I_1 (t, h; \epsilon; \mu) + I_1 (t, h; \epsbar; \mubar) \right], 
\nonumber 
\end{equation}
where $I_1 (t, h; \epsilon; \mu)$ is given in Eq. (\ref{eq:I1SB}). 

The variational equations [Eq. (\ref{eq:OPeq})] can be rewritten explicitly as follows. 
First, we notice that $\mu$ and $\mubar$ are related as $\mubar = \mu - \epsilon + \epsbar$. 
Remembering $\epsbar = \epsilon + 2 |\eta|$ [see just below Eq. (\ref{eq:EEbar})], this equation is reduced to 
\begin{equation}
\mubar = \mu + 2 |\eta|. 
\label{eq:OPeq_etabar}
\end{equation}
Further, substituting Eq. (\ref{eq:I1SB}) for $I_1$ in $f_0$ [Eq. (\ref{eq:f0})] and differentiating $f_0$ with respect to $\mu$, we get
\begin{eqnarray}
\frac{\del f_0}{\del \mu} 
&=& \frac{T_{\crm 0}}{2 \pi^2 {\xi_{ab}}^2 \xi_c} \, t h \,
\frac{\del I_1 (t, h; \epsilon; \mu)}{\del \mu} \nonumber \\
&=& \frac{T_{\crm 0}}{4 \pi^2 {\xi_{ab}}^2 \xi_c} \, t \nonumber \\ 
&& \times \int_0^{\sqrt{c - \epsilon}} \drm \widetilde{k} 
\Bigg[
\psi \left( \frac{\ktil^2 + \mu + c - \epsilon}{2 h} + \frac{1}{2} \right) \nonumber \\
&& \hspace{60pt} - \psi \left( \frac{\ktil^2 + \mu}{2 h} + \frac{1}{2} \right)
\Bigg]. 
\nonumber
\end{eqnarray}
Here, as in Appendix \ref{app:exp}, the digamma function $\psi (x) = \drm \ln \Gamma (x) / \drm x$ and a dimensionless wavenumber $\widetilde{k}$ were introduced. 
In the same way, the explicit form of $\del f_0 / \del \mubar$ is obtained. 
Lastly, substituting these explicit equations in Eq. (\ref{eq:OPeq}), we arrive at the following expression: 
\begin{equation}
\mu = \epsilon + \gamma \, t \left[ I_2 (t, h; \epsilon; \mu) + I_2 (t, h; \epsbar; \mubar) \right], 
\label{eq:OPeqTBexp}
\end{equation}
where $I_2$ is given in Eq. (\ref{eq:I2SB}). 
Here $\gamma$ is the fluctuation strength, defined by
\begin{equation}
\gamma = \frac{1}{4 \pi^2} \frac{\beta}{\alpha^2} 
\frac{\Tc}{{\xi_{ab}}^2 \xi_c} 
= \frac{1}{2 \pi^2} \frac{1}{\Delta C^\text{LEM} \, {\xi_{ab}}^2 \xi_c}, 
\label{eq:gamma}
\end{equation}
where $\Delta C^\text{LEM} = 2 \alpha^2 / (\beta \Tc)$ is the zero-field specific-heat jump at a mean-field level, determined only by the LEM. 

In this paragraph, we show the explicit form of the magnetization [Eq. (\ref{eq:M})]. 
By combining Eqs. (\ref{eq:f0}), (\ref{eq:M}), and (\ref{eq:I1SB}) with one another, we obtain 
\begin{eqnarray}
&& M_\text{dia} (t, h) \nonumber \\
&&= - \frac{2 \pi {\xi_{ab}}^2}{\phi_0} 
\frac{\Tc}{2 \pi^2 {\xi_{ab}}^2 \xi_c} \, t \nonumber \\
&& \hspace{20pt} \times \frac{\del [h \, I_1 (t, h; \epsilon; \mu) + h \, I_1 (t, h; \epsbar; \mubar)]}{\del h} \Bigg|_{\mu = \mu (t, h), \mubar = \mubar (t, h)} 
\nonumber \\
&&= - \frac{\Tc}{\pi \phi_0 \xi_c} \, t \nonumber \\ 
&& \hspace{20pt} \times \left[ I_3 (t, h; \epsilon; \mu) + I_3 (t, h; \epsbar; \mubar) \right] \big|_{\mu = \mu (t, h), \mubar = \mubar (t, h)}, 
\label{eq:MdiaTB}
\end{eqnarray}
where $I_3$ is given in Eq. (\ref{eq:I3SB}). 

\subsection{Cutoff-Independent Optimizing Equations and Magnetization Formula}
\label{app:cindepTB}

As in the single-band case, we show the asymptotic form of the optimizing equations [Eqs. (\ref{eq:OPeq_etabar}) and (\ref{eq:OPeqTBexp})] and magnetization formula [Eq. (\ref{eq:MdiaTB})] when the cutoff $c$ goes to infinity. 
As in Appendix \ref{app:cindepSB}, we adopt $\epsilon = T / \Tc - 1$ and $\epsbar = \epsilon + 2 |\eta|$ in this Appendix. 

By setting $h \rightarrow 0$ and $\mu = 0$ in Eq. (\ref{eq:OPeqTBexp}), we obtain the equation for determining the renormalized zero-field critical temperature $\TcTB (0)$ as 
\begin{eqnarray}
\frac{\TcTB (0)}{\Tc} - 1 &=& - 2 \gamtil^\text{(R)} \sqrt{c} - \gamtil^\text{(R)} \left( 2 \, \gamtil^\text{(R)} - \pi \sqrt{2 |\etatil|} \right) \no \\
&& \hspace{80pt} + \mathcal{O} \left( \frac{1}{\sqrt{c}} \right), 
\label{eq:TcRenTB}
\end{eqnarray}
where 
\begin{equation}
\left\{
\begin{array}{l}
\displaystyle \gamtil^\text{(R)} = \left( \frac{\pi}{2} + \ln 2 \right) \gamma^\text{(R)} = \left( \frac{\pi}{2} + \ln 2 \right) \frac{1}{4 \pi^2} \frac{\beta}{\alpha^2} \frac{\TcTB (0)}{{\xi_{ab}}^2 \xi_c} \vspace{5pt} \\
\displaystyle \etatil = \left( \frac{\pi}{2} + \ln 2 \right)^{-2} \eta. 
\end{array}
\right.
\end{equation}
In the above equation, $\gamtil^\text{(R)}$ is represented in terms of the renormalized zero-field critical temperature $\TcTB (0)$ and corresponds to Eq. (\ref{eq:gamtilSB}) in the single-band case. 

Since the asymptotic form of $I_2$ is already obtained in Eq. (\ref{eq:I2asymptSB}), we can find with the aid of Eq. (\ref{eq:TcRenTB}) that one of the optimizing equation (\ref{eq:OPeqTBexp}) takes the following form in $c \rightarrow \infty$ limit: 
\begin{eqnarray}
\mu &=& t^\text{(R)} - 1 + \gamma^\text{(R)} t^\text{(R)} \no \\
&& \times \bigg\{ \hspace{-3pt} - \hspace{-2pt} \int _0 ^\infty \hspace{-5pt} \drm \ktil \bigg[ \overline{\psi} \bigg( \frac{{\ktil}^2 + \mu}{2 h} + \frac{1}{2} \bigg) + \overline{\psi} \bigg( \frac{{\ktil}^2 + \mubar}{2 h} + \frac{1}{2} \bigg) \hspace{-2pt} \bigg] \no \\
&& \hspace{15pt} - \pi \left( \sqrt{\mu + h} + \sqrt{\mubar + h} \right) \no \\
&& \hspace{15pt} + \left( \frac{\pi}{2} + \ln 2 \right) \left[ 2 \, \gamtil^\text{(R)} \left( t^\text{(R)} - 1 \right) + \pi \sqrt{2 |\etatil|} \right] \bigg\}. \no \\
\label{eq:OPeqasymptTB}
\end{eqnarray}
Here, $\overline{\psi} (x)$ is defined in Eq. (\ref{eq:psibar}) and $t^\text{(R)} = T / \TcTB (0)$ is a dimensionless temperature expressed in terms of the renormalized critical temperature $\TcTB (0)$. 
Notice that Eq. (\ref{eq:OPeqasymptTB}) does not include either the cutoff $c$ or the bare critical temperature $\Tc$ but is represented in terms of the renormalized critical temperature $\TcTB (0)$. 
Combining Eqs. (\ref{eq:OPeq_etabar}) and (\ref{eq:OPeqasymptTB}), we now arrive at the final asymptotic expression of the set of optimizing equations. 

Regarding the magnetization formula [Eq. (\ref{eq:MdiaTB})] in $c \rightarrow \infty$ limit, the asymptotic forms of the gamma function $\Gamma (x)$ and the digamma function $\psi (x)$ used in Appendix \ref{app:cindepSB} immediately lead to the following expression: 
\begin{eqnarray}
M_\text{dia} &=& - \frac{\TcTB (0)}{\pi \phi_0 \xi_c} t^\text{(R)} \no \\
&& \times \int _0 ^\infty \drm \ktil \bigg[ \Upsilon \bigg( \frac{{\ktil}^2 + \mu}{2 h} + \frac{1}{2} \bigg) + \Upsilon \bigg( \frac{{\ktil}^2 + \mubar}{2 h} + \frac{1}{2} \bigg) \bigg], \no \\
\end{eqnarray}
where $\Upsilon (x)$ is already defined in Eq. (\ref{eq:Upsilon}). 
This formula clearly does not depend on either $c$ or $\Tc$ but is expressed in terms of $\TcTB (0)$. 

\subsection{Trial Free-Energy Density}
\label{app:fresTB}

In this appendix, we show how to derive Eq. (\ref{eq:ftri}). 
We recall that $f_\text{tri} = f_0 + \langle \Fcal_1 \rangle_0 / V$, and thus it is sufficient to write down how we treat $\langle \Fcal_1 \rangle_0$. 
First, $\Fcal_1$ is defined as 
\begin{eqnarray}
\Fcal_1 &=& \Fcal - \Fcal_0 \nonumber \\
&=& - \alpha (\mu - \epsilon) \sum_{N, q, k} |b_{Nqk}|^2 
- \alpha (\mubar - \epsbar) \sum_{\Nbar, \qbar, \kbar} |\bbar_{\Nbar \qbar \kbar}|^2 \nonumber \\
&& + \frac{\beta}{2} \int_V \drm^3 \rbm \left[ |\psi_1 (\rbm)|^4 + |\psi_2 (\rbm)|^4 \right]. 
\nonumber
\end{eqnarray}
Remember here that the indices $(N, q, k)$ and $(\Nbar, \qbar, \kbar)$ are attached to LEM and HEM, respectively, in order to recognize the difference in cutoff [Eq. (\ref{eq:cutoff})] for each mode. 
With Eqs. (\ref{eq:OPexpTB}) and (\ref{eq:a1a2}), then, we expand $\psi_1 (\rbm)$ and $\psi_2 (\rbm)$ as a linear combination of $b_{Nqk}$ and $\bbar_{\Nbar \qbar \kbar}$. 
The property of the Gaussian integral, such as $\langle b_{N q k}^* \bbar_{\hspace{1pt} \Nbar \qbar \kbar}^* \hspace{1pt} b_{N q k} \hspace{1pt} \bbar_{\hspace{1pt} \Nbar \qbar \kbar} \rangle_0 = \langle |b_{N q k}|^2 \rangle_0 \langle |\bbar_{\hspace{1pt} \Nbar \qbar \kbar}|^2 \rangle_0$, leads to
\begin{eqnarray}
&& \langle \Fcal_1 \rangle_0 \nonumber \\
&& = - \alpha (\mu - \epsilon) \sum_{N, q, k} \langle |b_{N q k}|^2 \rangle_0 
- \alpha (\mubar - \epsbar) \sum_{\Nbar, \qbar, \kbar} \langle |\bbar_{\hspace{1pt} \Nbar \qbar \kbar}|^2 \rangle_0 \nonumber \\
&& \hspace{10pt} + \frac{\beta}{2} \frac{1}{V} 
\bigg( 
\sum_{N, q, k} \langle |b_{N q k}|^2 \rangle_0
+ \sum_{\Nbar, \qbar, \kbar} \langle |\bbar_{\hspace{1pt} \Nbar \qbar \kbar}|^2 \rangle_0 \hspace{-2pt}
\bigg)^2. 
\label{eq:F1av}
\end{eqnarray}
From $f_0 = - (T / V) \ln \text{Tr}_{b, \bbar} \, \e^{- \Fcal_0 / T}$ and Eq. (\ref{eq:F0}), on the other hand, we see that 
\begin{equation}
\left\{
\begin{array}{l}
\displaystyle \frac{\del f_0}{\del \mu} 
= \frac{\alpha}{V} \sum_{N, q, k} \langle |b_{N q k}|^2 \rangle_0 \\
\displaystyle \frac{\del f_0}{\del \mubar} 
= \frac{\alpha}{V} \sum_{\Nbar, \qbar, \kbar} \langle |\bbar_{\hspace{1pt} \Nbar \qbar \kbar}|^2 \rangle_0. 
\end{array}
\right.
\label{eq:f0diff}
\end{equation}
Combining Eqs. (\ref{eq:F1av}) and (\ref{eq:f0diff}) together, we obtain
\begin{eqnarray}
\frac{\langle \Fcal_1 \rangle_0}{V}
&=& -(\mu - \epsilon) \frac{\del f_0}{\del \mu}
-(\mubar - \epsbar) \frac{\del f_0}{\del \mubar} \nonumber \\
&& + \frac{\beta}{2 \alpha^2} \left( \frac{\del f_0}{\del \mu} + \frac{\del f_0}{\del \mubar} \right)^2. 
\nonumber
\end{eqnarray}
From $f_\text{tri} = f_0 + \langle \Fcal_1 \rangle_0 / V$, therefore, we reach the expected equation [Eq. (\ref{eq:ftri})]: 
\begin{eqnarray}
f_\text{tri} 
&=& f_0
-(\mu - \epsilon) \frac{\del f_0}{\del \mu}
-(\overline{\mu} - \overline{\epsilon}) \frac{\del f_0}{\del \overline{\mu}} \nonumber \\
&& + \frac{\beta}{2 \alpha^2} \left( \frac{\del f_0}{\del \mu} + \frac{\del f_0}{\del \overline{\mu}} \right)^2. 
\nonumber
\end{eqnarray}

%%%%%%%%%%%%%%%%%

\section{$\textit{H}_\text{c2}$ Curve}
\label{app:TcH}

In this Appendix, we explain how we determine $T_\crm (H)$, which corresponds to the renormalized depairing field $H_\text{c2} (T)$ noted in 
Sec. \ref{sec:MM}. 
The `rough behavior' of magnetization is determined by low energy excitation modes (see Sec. \ref{sec:Tdep} and Fig. \ref{fig:M_Tdep_rough}). 
In our two-band case, therefore, the lowest-LL modes ($N = 0$) in LEM determine the `rough behavior.' 
In addition, we suppose that both of the high-LL ($N \geq 1$) modes in LEM and all LL modes in HEM will renormalize, i.e., lower, $T_\crm (H)$. 

In the variational equation [Eq. (\ref{eq:OPeq}), or explicitly, (\ref{eq:OPeqTBexp})], we can separate the effect of the lowest LL in LEM from other contribution as follows: 
\begin{eqnarray}
&& \mu + h \nonumber \\
&& = \left( \epsilon + h \right) \nonumber \\
&& \hspace{10pt} + \, 2 \, \gamma \, t h \, \frac{1}{\sqrt{\mu + h}} \arctan \sqrt{\frac{c - \epsilon}{\mu + h}} \nonumber \\
&& \hspace{10pt} + \Bigg\{ \gamma \, t \int_0^{\sqrt{c - \epsilon}} \drm \ktil 
\Bigg[ \psi \left( \frac{\ktil^2 + \mu + c - \epsilon}{2 h} + \frac{1}{2} \right) 
\nonumber \\
&& \hspace{90pt} - \psi \left( \frac{\ktil^2 + \mu}{2 h} + \frac{3}{2} \right) \Bigg] \Bigg\} \nonumber \\
&& \hspace{10pt} + \Bigg\{ \gamma \, t \int_0^{\sqrt{c - \epsbar}} \drm \ktil 
\Bigg[ \psi \left( \frac{\ktil^2 + \mubar + c - \epsbar}{2 h} + \frac{1}{2} \right) 
\nonumber \\
&& \hspace{90pt} - \psi \left( \frac{\ktil^2 + \mubar}{2 h} + \frac{1}{2} \right) \Bigg] \Bigg\}. 
\label{eq:OPeqmod}
\end{eqnarray}
Here the left hand side is a renormalized mass term of the lowest LL in LEM. 
We can simply interpret the right hand side of this equation as follows: 
The first term is the bare mass term of the lowest LL in LEM, which determines the rough value of $T_\crm (H)$. 
The second term represents the renormalization effect due to the interaction between fluctuations of the lowest LL in LEM, which generates the smooth behavior of magnetization around $T_\crm (H)$ (see Fig. \ref{fig:M_Tdep_rough}). 
The third term represents the contribution from high LLs in LEM, which is supposed to renormalize $T_\crm (H)$. 
The last term represents the contribution from all LLs in HEM, which is also supposed to renormalize $T_\crm (H)$. 

On the basis of the previous interpretation, we can suppose that the sum of the first, third, and last term in the right hand side of Eq. (\ref{eq:OPeqmod}) becomes zero when $T = T_\crm (H)$. 
Therefore, the equation determining $T_\crm (H)$ is given as 
\begin{eqnarray}
0 &=& \left( \epsilon + h \right) \nonumber \\
&& + \Bigg\{ \gamma \, t \int_0^{\sqrt{c - \epsilon}} \drm \ktil 
\Bigg[ \psi \left( \frac{\ktil^2 + \mu + c - \epsilon}{2 h} + \frac{1}{2} \right) 
\nonumber \\
&& \hspace{80pt} - \psi \left( \frac{\ktil^2 + \mu}{2 h} + \frac{3}{2} \right) \Bigg] \Bigg\} \nonumber \\
&& + \Bigg\{ \gamma \, t \int_0^{\sqrt{c - \epsbar}} \drm \ktil 
\Bigg[ \psi \left( \frac{\ktil^2 + \mubar + c - \epsbar}{2 h} + \frac{1}{2} \right) 
\nonumber \\
&& \hspace{80pt} - \psi \left( \frac{\ktil^2 + \mubar}{2 h} + \frac{1}{2} \right) \Bigg] \Bigg\}.
\end{eqnarray}

%%%%%%%%%%%%%%%%%

\section{Crossing of Magnetization Curves}
\label{app:crossing}

We comment on how to capture the temperature $T_\text{cross}$, at which the crossing of magnetization curves is seen in a substantial field range within our theoretical approach treating not only the lowest LL but also the higher LLs. 
The relation between $T_\text{cross}$ and phenomenological parameters has been theoretically investigated particularly in two-dimensional systems. On the other hand, the absence of a basic reason supporting the crossing behavior in three-dimensional case has been pointed out \cite{Welp,Sasik, Rosenstein}. Thus, in this Appendix, we focus on the following two-dimensional single-band system : 
\begin{eqnarray}
\widetilde{\Fcal} &=& \int_\Omega \drm^2 \rbm \, \alpha \bigg[ \epsilon |\psi|^2
+ {\xi_{ab}}^2 ( \bm{\Pi} \psi )^\dag ( \bm{\Pi} \psi ) \nonumber \\
&& \hspace{45pt} + \frac{\beta}{2 \alpha} |\psi|^4 \bigg], 
\end{eqnarray}
where $\Omega = L_x L_y$ means the system area, $\psi (\rbm)$ is expanded as $\psi (\rbm) = \sum_{N, q} \varphi_{N q} (x) {L_y}^{-1/2} \exp (\im q y) \, a_{N q}$, $\epsilon = \ln (T / \Tc) = \ln (t)$, and the other symbols are defined in the similar way to Sec. \ref{sec:methodSB}. 

In Ref. \onlinecite{Tesanovic}, the authors have considered only the lowest-LL fluctuation modes and found out an explicit relation between $T_\text{cross}$ and the bare transition temperature $\Tc$. 
In the following, however, we point out that their treatment seems invalid from the point of view of our approach. 
In our theoretical formalism, restricting oulselves to the lowest-LL fluctuation modes corresponds to separating the GL functional $\widetilde{\Fcal}$ in the form $\widetilde{\Fcal}^\text{LLL}_0 + \widetilde{\Fcal}_1$ and then applying the variational method explained in Sec. \ref{sec:methodSB}. 
Here $\widetilde{\Fcal}^\text{LLL}_0$ represents the part of the lowest-LL fluctuation with renormalized mass $\mu$: 
\begin{equation}
\widetilde{\Fcal}^\text{LLL}_0 = \sum_{q} \alpha \left( \mu + h \right) |a_{0q}|^2,  
\end{equation}
where $h = 2 \pi {\xi_{ab}}^2 \mu_0 H / \phi_0$ is a dimensionless magnetic field. 

In the same way as Sec. \ref{sec:methodSB}, we can obtain the magnetization $M_\text{dia}$ as a function of dimensionless temperature $t$ and magnetic field $h$: 
\begin{equation}
M_\text{dia} (t, h) = \frac{\Tc}{\phi_0} \, t \left\{ \ln \left[ \frac{\pi \Tc t}{\alpha (\mu + h)} \right] - \frac{h}{\mu + h} \right\} \Bigg|_{\mu = \mu (t,h)}. 
\label{eq:Mdia2D}
\end{equation}
$\mu (t, h)$ is the solution of the following variational equation: 
\begin{equation}
\mu = \epsilon + \gamma_\text{2D} \, \frac{t h}{\mu + h}. 
\label{eq:OPeq2D}
\end{equation}
Here $\gamma_\text{2D} = \beta \, \Tc / (\pi \alpha^2 {\xi_{ab}}^2)$ (which is usually $\ll 1$) represents the strength of the interaction between fluctuations. 

In the following, the explicit relation between $T_\text{cross}$ and $\Tc$ obtained in Ref. \onlinecite{Tesanovic} is discussed on the basis of Eqs. (\ref{eq:Mdia2D}) and (\ref{eq:OPeq2D}). 
\textit{Assuming} that the first term in Eq. (\ref{eq:Mdia2D}) is negligible, we may arrive at the following expression: $\del M_\text{dia} (t, h) / \del h \propto \mu (t, h) - h \, \del \mu (t, h) / \del h$. 
Then we can rewrite $\del \mu (t, h) / \del h$ as a function of $\mu (t, h)$ in terms of Eq. (\ref{eq:OPeq2D}), leading to $\del M_\text{dia} (t, h) / \del h \propto \mu (t, h)$. 
The crossing condition $\del M_\text{dia} (t, h) / \del h = 0$ is therefore equivalent to $\mu (t, h) = 0$. 
Lastly, by substituting $\mu = 0$ in Eq. (\ref{eq:OPeq2D}) and approximating $\epsilon = \ln (t) \simeq t - 1$, we obtain $t = (1 + \gamma_\text{2D})^{-1}$, or
\begin{equation}
T_\text{cross} = \frac{1}{1 + \gamma_\text{2D}} \, \Tc.
\label{eq:Tcross}
\end{equation}
The relation equivalent to this equation has been suggested in Ref. \onlinecite{Tesanovic}. 

Now we should consider the validity of the assumption that the first term in Eq. (\ref{eq:Mdia2D}) is negligible. 
When $\mu \sim -h$, the first term [$\sim \ln (\mu + h)$] indeed becomes much smaller than the second term [$\sim (\mu + h)^{-1}$]. 
The condition $\mu \sim -h$ will be achieved in the region where $\epsilon + h \simeq t + h - 1 \lesssim 0$ or $T / \Tc \lesssim 1 - h$, if $\gamma_\text{2D} (\ll 1)$ is not so large. 
On the other hand, the resulting crossing temperature [Eq. (\ref{eq:Tcross})] satisfies $T_\text{cross} / \Tc \simeq 1 - \gamma_\text{2D}$. 
When $h = 0.1$, for example, $T_\text{cross} / \Tc > 1 - h$, and therefore the assumption that the first term in Eq. (\ref{eq:Mdia2D}) is negligible is not justified. 

On the contrary to the previous discussion, we have tried calculating the crossing temperature $T_\text{cross}$, treating high LLs as well as the lowest LL in the same way as in Sec. \ref{sec:methodSB}. 
As a result, though we cannot analytically find out the field-independent crossing temperature, we numerically obtain a temperature which can, as in Fig. \ref{fig:M_Tdep_highh}, be identified with a crossing point at least in a substantial field range. 
In conclusion, we argue that the procedure of arriving at the above expression [Eq. (\ref{eq:Tcross})] will not be justified in moderately high fields, and that nevertheless, our approach including not only the lowest-LL modes but also higher-LL ones can approximately produce a crossing behavior of the magnetization curves. 

%%%%%%%%%%%%%%%%%

%%%%%%%%%%%%%%%%%


\begin{thebibliography}{99}
\bibitem{eta0lowT} When $\eta \simeq 0$, the HEM in addition to LEM can significantly contribute to the zero-field specific-heat jump $\Delta C$. 
This fact makes ambiguous the way to choose $\gamma$ [see Eq. (\ref{eq:gamma})] when we compare the two-band case with the single-band case. 
Thus, inside and below the critical region, where $\gamma$ plays an important role in determining the value of magnetization, a simple comparison between the two-band and single-band cases is difficult when $\eta \simeq 0$.
\bibitem{material} Material parameters are set as $\xi_{ab} = 5.5 \, \text{nm}$, $\xi_c = 1.5  \, \text{nm}$, and $\Tc = 8.5 \, \text{K}$.\cite{Kasahara} 
Other parameters are set as $c = 0.6$, $\gamma = 0.001$, and $\eta = 0.05$, i.e., the same values as used in Fig. \ref{fig:M_Tdep_rough} - \ref{fig:M_hdep}. 
Change of $c$ within $\mathcal{O}(1)$ does not significantly affect the result; 
Change in $\gamma$ is also trivial since the compared temperature region is largely outside the critical region; 
Though $\eta$ is important for the resulting magnetization, we find $|\chi_\text{dia}^\text{exp}| > |\chi_\text{dia}|$ even in the extreme case where $\eta \rightarrow 0$ (see Sec. \ref{sec:etadep}).
\end{thebibliography}
\end{document}